\documentclass[english,a4paper,aps,twocolumn,amsmath,amssymb,longbibliography]{revtex4-1}
\usepackage{graphicx,color}
\usepackage{bm}

\newcommand{\pd}[2]{\frac{\partial #1}{\partial #2}}
\newcommand{\fd}[2]{\frac{\delta #1}{\delta #2}}
\newcommand{\mean}[1]{\langle #1\rangle}
\newcommand{\Int}[1]{\int\text{d}#1\;}
\newcommand{\IInt}[3]{\int_{#2}^{#3}\text{d}#1\;}
\renewcommand{\vec}[1]{\mathbf #1}
\DeclareMathOperator{\tr}{tr}

\newcommand{\al}{\alpha}
\newcommand{\gam}{\gamma}

\newcommand{\kap}{\kappa}
\newcommand{\lam}{\lambda}
\newcommand{\Lam}{\Lambda}

\newcommand{\sig}{\sigma}
\newcommand{\x}{\vec r}
\newcommand{\nois}{\bm\xi}
\newcommand{\msig}{\bm\sig}
\newcommand{\Dr}{D_\text{r}}
\newcommand{\tx}{\tau_\text{r}}
\newcommand{\id}{\mathbf 1}
\newcommand{\kT}{k_\text{B}T}

\begin{document}

\title{Collective forces in scalar active matter}

\author{Thomas Speck}
\affiliation{Institut f\"ur Physik, Johannes Gutenberg-Universit\"at Mainz, Staudingerweg 7-9, 55128 Mainz, Germany}

\begin{abstract}
  Large-scale collective behavior in suspensions of many particles can be understood from the balance of statistical forces emerging beyond the direct microscopic particle interactions. Here we review some aspects of the collective forces that can arise in suspensions of self-propelled active Brownian particles: wall forces under confinement, interfacial forces, and forces on immersed bodies mediated by the suspension. Even for non-aligning active particles, these forces are intimately related to a non-uniform polarization of particle orientations induced by walls and bodies, or inhomogeneous density profiles. We conclude by pointing out future directions and promising areas for the application of collective forces in synthetic active matter, as well as their role in living active matter.
\end{abstract}

\maketitle


\section{Introduction}

Over the last decade, ``active matter'' has moved more and more into the focus of several disciplines, providing a paradigm for a range of driven systems belonging to soft and living matter~\cite{roadmap}. These systems are composed of many interacting but autonomous ``entities'' that move diffusively but, in contrast to passive diffusion, are characterized by a persistence of their motion, \emph{i.e.}, displacements remain correlated over some finite time. Such a persistence breaks \emph{detailed balance} and is only possible if available (free) energy is dissipated. While active matter has kindled this interest mainly due to the wealth of the emergent collective behavior, intimately related are non-equilibrium ``statistical'' collective forces, the physics of which has received comparatively little attention~\cite{basu15a}.

\begin{figure}[b!]
  \centering
  \includegraphics{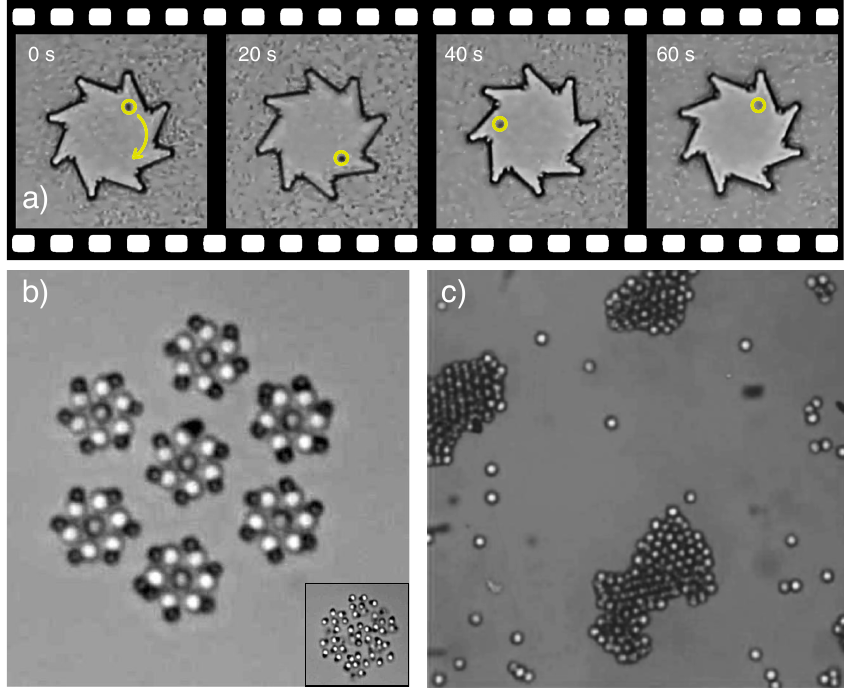}
  \caption{Manifestations of collective forces in active matter. (a)~Gear (diameter $48\;\mu$m) rotating clockwise in a bath of bacteria [reproduced from Ref.~\citenum{leon10}]. (b)~Self-assembled colloidal spinners [reproduced from Ref.~\citenum{aubret18}]. (c)~Coexistence of self-propelled colloidal Janus particles stabilized by active interfacial forces [reproduced from Ref.~\citenum{butt13}].}
  \label{fig:sys}
\end{figure}

Collective forces can appear as gradients of thermodynamic potentials generating fluxes that bring the system back to equilibrium, but also Casimir-like forces due to fluctuations~\cite{kardar99,hert08}. The balance of collective interfacial forces underlies the many inhomogeneous states observed in soft matter systems, \emph{e.g.}, emulsions of water and oil. Freed from the constraints of detailed balance, related phenomena in active matter are more faceted and also allow for steady currents as observed, \emph{inter alia}, in the spontaneous coherent flow of a suspension of microtubules and kinesin motors~\cite{wu17}. Figure~\ref{fig:sys}a shows an example for collective torques in active suspensions: the directed rotation of microgears because of their asymmetric shape~\cite{leon10,sokolov10}. Such torques (and their coupling) have been exploited for the hierarchical organization of spinners into a lattice~\cite{aubret18}, cf. Fig.~\ref{fig:sys}b. Figure~\ref{fig:sys}c shows an example for the clustering (and phase separation) of repulsive Janus particles in the absence of coherent forces~\cite{butt13}.

Theoretical approaches classify active matter as scalar, polar, or nematic depending on the rank of the (bulk) order parameter that is to be modeled: scalar density, vectorial polarization, or tensorial nematic order. Moreover, coupling to an explicit fluid obeying momentum conservation is considered ``wet'' while its absence gives rise to the notion of ``dry'' active matter~\cite{chate19}.

The purpose of the present manuscript is to give an overview of the (chiefly theoretical) work concerned with collective forces arising in non-aligning scalar active matter and to summarize the current understanding. To expose and discuss basic physical principles, we will focus on the simplest class of models neglecting long-range hydrodynamic and phoretic forces, but also explicit alignment torques. We will adapt a soft/liquid matter perspective based on the stresses within a suspension of interacting active particles.

Although active matter is a rapidly progressing field there is already an impressive number of reviews covering different aspects. To name a few (this is not an exhaustive list): swimming and hydrodynamics at the microscale~\cite{laug09,elge15}, active (mainly colloidal) particles~\cite{bech16,zott16}, and the theory of motility-induced phase separation~\cite{cate15}. Regarding living matter, notable reviews are available on the theory of active gels~\cite{pros15} and on the cell as an active material~\cite{needleman17}.


\section{Theory of Active Brownian particles}

We focus on active Brownian particles (ABPs), the arguably simplest model for $N$ identical self-propelled and interacting particles moving in $d$ dimensions. Every particle is described by its position $\x_i$ and unit orientation $\vec e_i$. Throughout we assume that friction is large, \emph{i.e.}, the coupling to the solvent (or substrate) is such that the momentum relaxes on timescales much shorter than those of interest. The overdamped equations of motion for each particle can then be written
\begin{equation}
  \dot\x_i = v_0\vec e_i + \mu_0\vec F_i + \nois_i,
  \label{eq:lang}
\end{equation}
where $v_0$ is the bare propulsion speed and $\vec F_i=-\nabla_iU$ is the force stemming from the potential energy $U(\{\x_i\})$, which only depends on particle positions but not orientations. We include translational noise $\nois_i$, which we assume to be Gaussian with correlations determined by $D_0$. The fluctuation-dissipation theorem~\cite{kubo66} relates the diffusion coefficient $D_0=\mu_0\kT$ to the bare mobility $\mu_0$ through the temperature $T$ (with Boltzmann's constant $k_\text{B}$). We will keep this relation, in case mobility and diffusion are independent parameters it simply defines an effective temperature. The orientations are not fixed but undergo rotational diffusion with diffusion coefficient $\Dr$ and correlation time $\tx=[(d-1)\Dr]^{-1}$. While no-slip boundary conditions with a solvent restrict $\Dr=3D_0/a^2$ for spherical particles with diameter $a$, in general $\Dr$ is a free parameter. The limiting cases are $\tx\to0$, for which the propulsion term becomes uncorrelated noise, while $\tx\to\infty$ corresponds to ``ballistic active motion''~\cite{bruss18,reich18}. For a free particle, the combination of directed motion and orientational diffusion creates trajectories that are characterized by a persistence length $\ell_\text{p}=v_0\tx$~\cite{hows07}.

Equivalently to considering the $N$ coupled stochastic equations of motion~\eqref{eq:lang}, we can consider the joint probability distribution $\psi_N(\{\x_i,\vec e_i\};t)$ of all positions and orientations. It obeys the evolution equation
\begin{equation}
  \partial_t\psi_N = -\sum_{i=1}^N\nabla_i\cdot\mathcal J_i + \Dr\sum_{i=1}^N\Delta_i\psi_N,
  \label{eq:fp:N}
\end{equation}
with probability currents $\mathcal J_i=v_0\vec e_i\psi_N+\mu_0\vec F_i\psi_N-D_0\nabla_i\psi_N$, where $\Delta_i$ is the Laplace operator on the $d$-dimensional unit sphere acting on $\vec e_i$. For $v_0=0$, the stationary distribution is given by the Boltzmann weight $\psi_N\sim e^{-U/\kT}$ with uniformly distributed orientations. For $v_0\neq0$ breaking detailed balance no closed solution exists. Already the stationary distribution for a single particle in the presence of walls is non-trivial~\cite{wagner17,hermann18}. Some results for the time-dependent distribution have been obtained for limiting cases~\cite{basu18,basu19}.

\begin{figure}[t]
  \centering
  \includegraphics{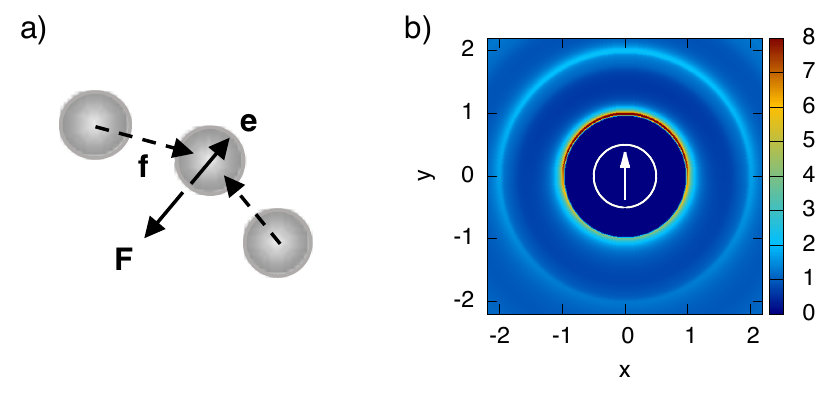}
  \caption{Tagged particle. (a)~Direct pair forces $\vec f$ of other particles exerted on the tagged particle with orientation $\vec e$ (arrow). Averaging over the environment (for a fixed tagged particle) yields the conditional force $\vec F\propto-\vec e$. (b)~Pair distribution function $g(\x)$ from simulations of hard discs. From the perspective of each particle there is an increased density of particles in font, and a reduced density in the back. Reproduced from Ref.~\citenum{bial13}.}
  \label{fig:forces}
\end{figure}

We are interested in collective forces on lengths larger than the particle diameter. To proceed it is crucial to split forces $\vec F_i=\vec F^\text{ex}(\x_i)+\vec F^\text{int}_i$ into external one-body forces $\vec F^\text{ex}(\x)$ due to, \emph{e.g.}, walls, and forces $\vec F^\text{int}_i=\sum_{j\neq i}\vec f(\x_i-\x_j)$ due to the pairwise particle interactions $\vec f=-\nabla u$ with pair potential $u(r)$. These interaction forces can be related to the divergence of the symmetric Irving-Kirkwood stress tensor~\cite{irving50},
\begin{equation}
  \nabla\cdot\msig_\text{IK} = \left\langle \sum_{i=1}^N \vec F^\text{int}_i\delta(\x-\x_i) \right\rangle = \mean{\vec F}_\x,
  \label{eq:IK}
\end{equation}
whereby the brackets $\mean{\cdot}$ denote the average over the joint probability $\psi_N$. For pairwise interactions, we define the conditional force
\begin{equation}
  \vec F(\x,\vec e;t) = \Int{^d\x'} \vec f(\x-\x') \psi_2(\x'|\x,\vec e;t)
  \label{eq:F}
\end{equation}
on a tagged particle [Fig.~\ref{fig:forces}a], where $\psi_2(\x'|\x,\vec e;t)$ is the conditional density to find a second particle at position $\x'$ given that the tagged particle with orientation $\vec e$ resides at $\x$. The force Eq.~\eqref{eq:IK} is then obtained as average $\mean{\cdot}_\x$ over the tagged particle orientation.

Integrating the evolution equation~\eqref{eq:fp:N} over all degrees of freedom except the position $\x$ of a single tagged particle yields the continuity equation $\partial_t\rho+\nabla\cdot\vec j=0$ with density $\rho(\x;t)=\mean{\sum_i\delta(\x-\x_i)}$ and particle current
\begin{equation}
  \vec j = v_0\vec p + \mu_0\nabla\cdot\msig_\text{IK} + \mu_0\vec F^\text{ex}\rho - D_0\nabla\rho.
  \label{eq:j}
\end{equation}
Within our model, this is an exact result. For $v_0=0$ and vanishing current $\vec j=0$, it reduces to the hydrostatic equilibrium condition, $\nabla\cdot(-\kT\rho\id+\msig_\text{IK})+\vec F^\text{ex}\rho=0$. For self-propelled particles, the current couples to the local polarization $\vec p(\x;t)=\mean{\sum_i\vec e_i\delta(\x-\x_i)}$.

Besides the non-Boltzmann stationary distribution, this polarization is what sets active particles apart from passive systems as it directly enters the balance of forces Eq.~\eqref{eq:j}. An evolution equation can be obtained by multiplying Eq.~\eqref{eq:fp:N} by the orientation of the tagged particle $\vec e$ followed again by integration,
\begin{equation}
  \partial_t\vec p = -\nabla\cdot\left[\frac{v_0}{d}\rho\id+v_0\vec Q+\mu_0\vec C+\mu_0\vec p\vec F^\text{ex}-D_0\nabla\vec p\right] - \frac{\vec p}{\tx}.
  \label{eq:p}
\end{equation}
It couples to two tensor fields, the symmetric and traceless nematic tensor $\vec Q(\x;t)=\mean{(\vec e\vec e-\id/d)}_\x$ and the tensor
\begin{equation}
  \vec C(\x;t) = \left\langle\sum_{i=1}^N\vec e_i\vec F^\text{int}_i\delta(\x-\x_i)\right\rangle = \mean{\vec e\vec F}_\x
\end{equation}
capturing the correlations between orientations and forces [throughout, we write direct products as $(\vec e\vec e)_{kl}=e_ke_l$]. Clearly, this approach creates an infinite hierarchy of coupled evolution equations for the moments of the orientation ($\rho$, $\vec p$, $\vec Q$, \dots)~\cite{saint15}. To obtain a useful representation, we need to close the hierarchy.

In a homogeneous system with density $\rho=\bar\rho$ in steady state, the conditional density $\psi_2=\rho g(\x'-\x|\vec e)$ only depends on the separation $\x-\x'$ and, moreover, is axisymmetric with respect to the orientation $\vec e$ of the tagged particle. This implies that the conditional force can be written $\vec F=-\zeta\rho\vec e$ with a coefficient $\zeta=\zeta(\rho,v_0)$ that depends on the pair potential $u(r)$ and the shape of the pair distribution $g(\x)$. In an isotropic system $\zeta=0$, while for self-propelled particles, on average, there will be more particles in front than in the back (Fig.~\ref{fig:forces}b), leading to a non-vanishing conditional force $\vec F$ opposing the directed motion. We thus obtain the \emph{force closure}
\begin{equation}
  \mean{\vec F}_\x = -\zeta\rho\vec p, \qquad
  \vec C = \mean{\vec e\vec F}_\x = -\zeta\rho\vec Q-\zeta\rho^2\id/d,
  \label{eq:closure}
\end{equation}
which implies that the interparticle interactions can be represented through the density-dependent speed $v(\rho)=v_0-\mu_0\zeta\rho$. While strictly valid only for homogeneous systems, the force closure has been used for inhomogeneous systems as long as the density varies slowly (much slower than the range of the direct forces $\vec f$)~\cite{spec15}. At this stage the coefficient $\zeta$ is an input capturing the two-body statistics. For (almost) hard discs a density-independent coefficient $\zeta\simeq v_0/(\mu_0\rho_0)$ has been determined in simulations~\cite{sten13} (with density $\rho_0$ at which the system ``jams''). Approximate expressions for the pair distribution $g(\x)$ have been derived for hard discs~\cite{hart18}, with exact results available in infinite dimensions~\cite{pirey19}.

At the next level of the hierarchy, we obtain an evolution equation for $\vec Q$. The rotational diffusion leads to a term $-2d\Dr\vec Q$ with time constant $(2d\Dr)^{-1}<\tx$, which implies that the nematic order decays faster than the polarization. This might justify to set $\vec Q\approx 0$ in order to close the hierarchy, which seems to be a reasonable approximation close to the uniform state. To lowest non-vanishing order with $\vec F^\text{ex}=0$ one obtains~\cite{bert06}
\begin{equation}
  Q_{kl} \approx -\al_d\tx\left[\partial_k(vp_l)+\partial_l(vp_k)-\frac{2}{d}\nabla\cdot(v\vec p)\delta_{kl}\right]
  \label{eq:Q}
\end{equation}
in cartesian coordinates with dimensionless coefficient $\al_d$ (in two dimensions $\al_2=\tfrac{1}{16}$). It expresses the nematic tensor entirely through derivatives of the polarization and thus also closes the hierarchy. Note that in the homogeneous state polarization $\vec p=0$ and nematic order $\vec Q=0$ vanish, which makes ABPs the paradigm for scalar active matter.

Another route to derive similar hydrodynamic equations for the lowest moments is to start from the underdamped Langevin equations and to perform coarse-graining~\cite{epstein19} or a systematic timescale separation~\cite{steff17}. An alternative way to interpret Eq.~\eqref{eq:lang}, which we will not discuss further, is to eliminate the orientations leading to stochastic differential equations for the positions that now involve colored translational noises with correlation time $\tx$. In certain limits one can derive a Markovian evolution equation for the joint distribution of the positions with effective forces and diffusion coefficients that now depend on the speed $v_0$ and the correlation time $\tx$~\cite{fara15,magg15,marc15,rein16a}.

\section{Collective forces}

\subsection{Active stress}

We now focus on the steady state. With $\partial_t\vec p=0$ we rearrange Eq.~\eqref{eq:p} to $v_0\vec p=\mu_0\nabla\cdot\msig_\text{A}$ with active stress
\begin{equation}
  \msig_\text{A} = -v_0\tx\left[\frac{v_0}{d\mu_0}\rho\id+\frac{v_0}{\mu_0}\vec Q+\vec C+\vec p\vec F^\text{ex}-\kT\nabla\vec p\right].
  \label{eq:sig:a}
\end{equation}
Note that this active stress has an isotropic component that scales as $v_0^2$, whereas the deviatoric stress is determined entirely by the polarization. Eliminating the polarization in Eq.~\eqref{eq:j}, we obtain the force balance
\begin{equation}
  \vec j/\mu_0 = \nabla\cdot\msig + \vec F^\text{ex}\rho
  \label{eq:j:sig}
\end{equation}
with total stress tensor $\msig=-\kT\rho\id+\msig_\text{IK}+\msig_\text{A}$.

\subsection{Wall forces and pressure}

\begin{figure}[t]
  \centering
  \includegraphics{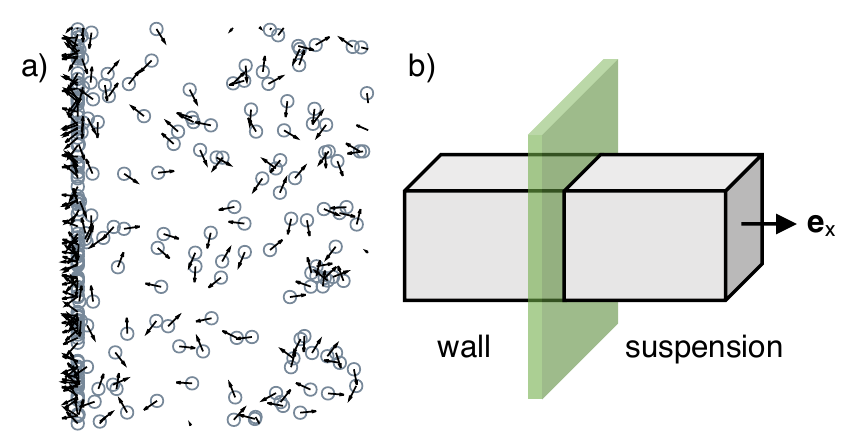}
  \caption{Wall forces in confined suspensions. (a)~Accumulation of non-interacting (ideal) ABPs in front of a wall. Note that trapped particles point into the wall. [Reproduced from Ref.~\citenum{spec16a}]. (b) Sketched is an integration volume $V$ (gray). Using the divergence theorem, the force on the wall (green) can be converted to a surface integral to which only the dark gray surface within the suspension contributes.}
  \label{fig:wall}
\end{figure}

Confined liquids exert a pressure on the walls of their container. In contrast to passive particles, active particles aggregate at walls due to the blocked persistent motion, increasing the local density and inducing a non-vanishing local polarization (trapped particles point into the wall, cf. Fig.~\ref{fig:wall}a). Initial computer simulations show that the trapped particles indeed exert an additional force that can be interpreted as an elevated pressure~\cite{mall14,yang14}.

To calculate this pressure, let us consider a piece of flat wall that exerts a force $\vec F^\text{ex}$ on the active particles (we orient the coordinate system so that the wall normal is $\vec e_x$). Newton's third law demands that this force also acts onto the wall so that the total wall force becomes $\vec F^\text{wall}=-\IInt{^d\x}{V}{}\vec F^\text{ex}\rho$. The integration volume $V$ is a rectangular prism with cross section $A$ completely enclosing the region in which $\vec F^\text{ex}\neq 0$, cf. Fig.~\ref{fig:wall}b. We assume that the system is translationally invariant parallel to the wall, which precludes a parallel particle current while the wall itself precludes a current along the direction of the wall normal so that $\vec j=0$. Density and polarization can and do vary along the $x$-direction.

The force balance Eq.~\eqref{eq:j:sig} together with the divergence theorem yields $\vec F^\text{wall}=\oint_{\partial V}\text{d}^{d-1}\x\;\vec n\cdot\msig$ (with normal vector $\vec n$). Due to the translational invariance, the lateral integration surfaces of the integration volume $V$ cancel each other while the surface within the wall does not contribute. We are thus left with the stress at the top surface within the active suspension. The force on the wall is independent of the position $x$ (as long as it is outside the wall with $\vec F^\text{ex}=0$). We thus push the top surface to a distance where there is no influence of the wall on the suspension anymore, \emph{i.e.}, it is homogeneous and isotropic with uniform density $\bar\rho$ and vanishing $\vec p=0$, $\vec Q=0$. We obtain the wall force $\vec F^\text{wall}=-pA\vec e_x$ with (mechanical) pressure
\begin{equation}
  p = \kT\bar\rho + p_\text{IK} + \frac{v_0\tx v(\bar\rho)}{d\mu_0}\bar\rho
  \label{eq:pres}
\end{equation}
using the (in the homogeneous suspension exact) force closure Eq.~\eqref{eq:closure}. The pressure in the suspension due to the interaction forces is $p_\text{IK}=-\tr\msig_\text{IK}/d$. The result Eq.~\eqref{eq:pres} has been obtained following different routes: treating propulsion as a swim force~\cite{taka14,taka14a,spec16}, through correlation functions~\cite{solo15}, and deriving the pressure from the virial~\cite{wink15,spec16a,levi17,das19}. To calculate the pressure one can also estimate the excess density within the interaction range of walls~\cite{yan15a,duzgun18}, see Ref.~\citenum{ezhi15} for run-and-tumble dynamics. In experiments, an active pressure has been measured from sedimentation profiles~\cite{gino15} and from the deformation of a flexible membrane~\cite{juno17}.

Using the divergence theorem has allowed us to connect the mechanical pressure onto a wall with the homogeneous state of the suspension characterized by a few quantities ($\bar\rho$, $T$, $v_0$) independent of the wall interactions. The situation changes fundamentally if one considers walls that exert torques on the particles, \emph{i.e.}, the external potential $U^\text{ex}(\x,\vec e)$ is a function of both position and orientation~\cite{solo15a}. In this case the external potential cannot be eliminated and enters the pressure.

While we have determined the force onto a wall from properties of the uniform suspension, what is the \emph{local} pressure within the suspension? To this end, we write the force balance Eq.~\eqref{eq:j:sig} as
\begin{equation}
  0 = \frac{v_0}{\mu_0}\vec p + \nabla\cdot(-\kT\rho\id+\msig_\text{IK}) + \vec F^\text{ex}\rho
  \label{eq:j:p}
\end{equation}
eliminating the active stress in favor of the polarization. Suppose we instantly insert a wall into a uniform suspension. Right after inserting the wall but before the trapping and accumulation of particles at the wall we have $\rho=\bar\rho$ and $\vec p=0$ everywhere so that the instantaneous pressure reads $p_\text{loc}=\kT\bar\rho+p_\text{IK}$~\cite{spec16a,omar19}. This pressure will then increase and reach Eq.~\eqref{eq:pres} in the steady state.

\subsection{Interfacial forces}

\subsubsection{Passive liquids and suspensions}

We now consider inhomogeneous systems in the absence of external forces, $\vec F^\text{ex}=0$. In equilibrium ($v_0=0$), inhomogeneity implies that there are two phases that coexist and which are separated by an interface. The fact that there is a stable density gradient requires forces within the interface that balance the diffusive flux $\propto\nabla\rho$. We orient the coordinate system so that the normal of the (on average flat) interface points along $x$. Moreover, we assume that the system is invariant with respect to rotation about the axis given by the normal so that the in-plane stress components $\sig_{yy}=\sig_{zz}=\sig_\parallel(x)$ are equal. From the force balance $\nabla\cdot\msig=0$ we can conclude that the normal component $\sig_\perp$ has to be constant throughout the system. The interfacial tension $\gam$ between the two coexisting phases can be determined through two different routes, which in equilibrium yield the same result.

The mechanical route following Kirkwood and Buff~\cite{kirk49} starts from the work $\delta w=V\tr(\msig\cdot\vec h)$ necessary to deform a subsystem by moving particles according to $\x_k\to(\id+\vec h)\cdot\x_k$. The matrix $\vec h$ has only diagonal entries, $h_{xx}=h_\perp$ and $h_{yy}=h_{zz}=h_\parallel$. We slice the system into bins of width $\delta x$ along the normal. The total work then becomes $w=\sum_kV[\sig_\perp h_\perp+(d-1)\sig_\parallel(x_k)h_\parallel)]$ summing over bins. The bin volume $V=A\delta x$ and area $A$ change as $\delta V=V[h_\perp+(d-1)h_\parallel]$ and $\delta A=A(d-1)h_\parallel$. Keeping the volume fixed ($\delta V=0$) yields $w=\gam\delta A$ with
\begin{equation}
  \gam = \IInt{x}{-\infty}{+\infty} [\sig_\parallel(x)-\sig_\perp]
  \label{eq:gam}
\end{equation}
after taking the limit $\delta x\to0$. On the other hand, the thermodynamic route asks for the change of the grand potential $\Omega$ under the same geometric transformation. Working out how the pair potential changes~\cite{evan79} yields $\delta\Omega=\gam\delta A$ with the same expression Eq.~\eqref{eq:gam} as the mechanical route. This is of course to be expected since in equilibrium $\delta\Omega=\delta w$ is the reversible work to change the interfacial area.

Detailed balance guarantees that the system is relaxing towards thermal equilibrium governed by the Boltzmann distribution and a free energy. In the vicinity of a critical point at ($T_\text{c}$,$\rho_\text{c}$), this free energy can be expanded into the well-known Ginzburg-Landau expression~\cite{hohe15}
\begin{equation}
  \label{eq:gl}
  \mathcal F[\rho] = \int \text{d}^d\x\; \left\{ \frac{\kappa}{2}|\nabla\rho|^2 + f_0(\rho) \right\}
\end{equation}
with free energy density $f_0(\rho)\approx a(\rho-\rho_\text{c})^2+b(\rho-\rho_\text{c})^4$, the minima of which determine the coexisting densities $\rho_\pm$. The coefficient $\kappa>0$ determines the cost of interfaces yielding the interfacial tension $\gam=\Int{x}\kap|\partial_x\rho|^2$~\cite{evan79}. Note that the coexisting densities are independent of $\kap$.

\subsubsection{Active suspensions}
\label{sec:direct}

\begin{figure}[b!]
  \centering
  \includegraphics{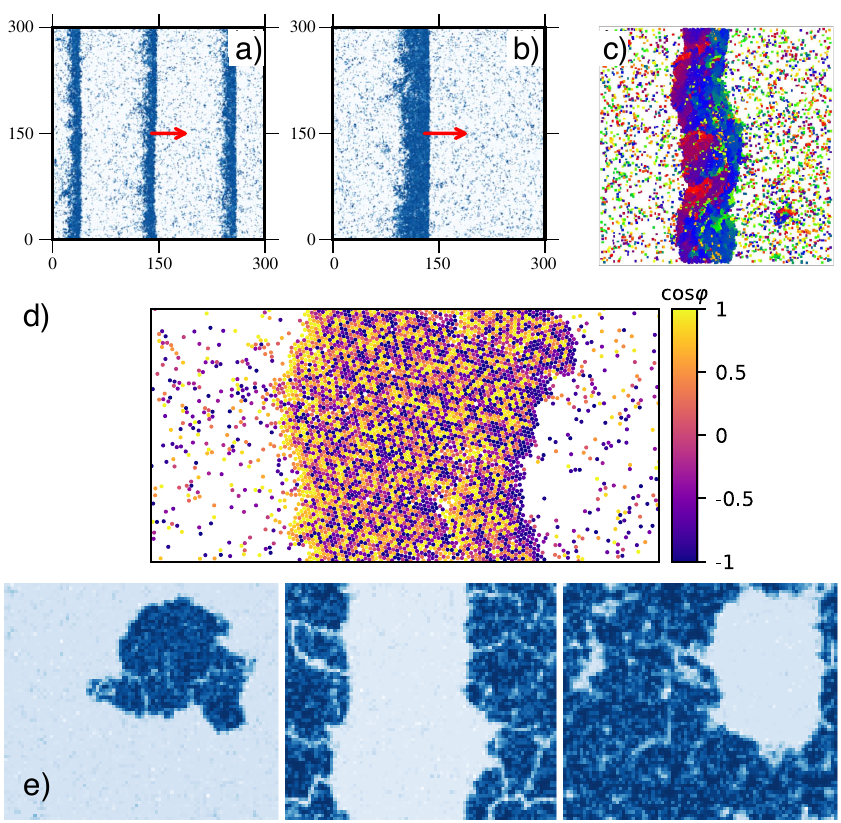}
  \caption{Coexistence in active matter. (a,b)~Dense traveling bands in the Vicsek model [Reproduced from Ref.~\citenum{caussin14}]. (c)~Coexistence of a polar band with a gas [Reproduced from Ref.~\citenum{weitz15}]. (d)~Simulation snapshot of phase-separated ABPs. The color indicates the orientation, note the polarization within the interfaces where particles point into the dense domain. (e)~Finite-size transitions in ABPs as a function of global density (increasing from left to right) [Reproduced from Ref.~\citenum{bial15}].}
  \label{fig:coex}
\end{figure}

Active matter ($v_0\neq 0$) breaks detailed balance, and, strictly speaking, no thermodynamic potential exists. Still, many active systems show qualitatively very similar behavior, in particular the coexistence of two ``phases'' separated by an interfacial region. In Fig.~\ref{fig:coex} we show three numerical examples of inhomogeneous active systems: traveling bands in the Vicsek model (alignment without volume exclusion, Fig.~\ref{fig:coex}a,b), the coexistence of a polar band with an active gas in polar rods (alignment with volume exclusion, Fig.~\ref{fig:coex}c), and the coexistence of dilute and dense regions of ABPs (no alignment but volume exclusion, Fig.~\ref{fig:coex}d,e).

Again focusing on ABPs (Fig.~\ref{fig:coex}d), their phase diagram resembles that of passive liquid-gas coexistence with speed $v_0$ taking the role of inverse temperature~\cite{redn13,fily14,bial14,spec15,digregorio18}. In particular, the coexistence region is terminated by a critical point~\cite{sieb18} below which the suspension remains homogeneous. In the two-phase region ($v_0>v_\text{c}$), we observe that ABPs undergo a series of transitions as the global density is increased, where the morphology of the dense domain changes from droplet to a slab to a ``bubble'' (Fig.~\ref{fig:coex}e). This is well understood in passive liquid-gas phase separation as finite-size transitions~\cite{bind03,schr09}: minimizing the length of the interface (in a finite box) under the constraint of the lever rule yields exactly these shapes. Apparently an effective interfacial tension also governs the coexistence of ABPs, although it does not arise from a free energy.

The interfacial tension Eq.~\eqref{eq:gam} follows from a purely mechanical argument based on the anisotropy of stresses due to the presence of the interface. Interestingly, evaluating this expression for simulations of repulsive ABPs yields a negative and surprisingly large value ($|\gam|\sim10^3$ for hard discs compared to $\gam\sim1$ for passive liquid-gas coexistence)~\cite{bial15}. This result challenges our intuition (which, however, is shaped by the idea of a free energy): a negative tension would imply that the system could reduce its free energy by enlarging the interface, which would lead to a proliferation of interfacial area and eventually to a homogeneous system. Such an argument stands in stark contrast to the observed stable phase separation of ABPs.

To reconcile a negative interfacial tension with stable phase coexistence it is important to recognize the role played by the active stress $\msig_\text{A}$. Even though there is no explicit alignment of ABPs, the polarization within the interface is non-zero. This can be understood easily: particles arriving from the gas point into the dense phase and are trapped at the interface due to the persistence of their orientations. On the other hand, particles pointing outwards can quickly leave the interfacial region, which thus leads to an excess of particles oriented towards the dense phase. This means that the nematic tensor now contributes a term $-(v_0\tx v/\mu_0)\vec Q$ to the active stress [Eq.~\eqref{eq:sig:a}], where we have used the force closure. From Eq.~\eqref{eq:Q} one finds $Q_\perp\approx-\al_d[2(d-1)/d]\tx\partial_x(vp_x)$. With $Q_\parallel=-Q_\perp/(d-1)$ and using $vp_x\approx D_0\partial_x\rho$ from the force balance Eq.~\eqref{eq:j}, we obtain (after some integrations by parts)
\begin{equation}
  \gam_\text{A} \approx \IInt{x}{-\infty}{+\infty}\kap_\text{A}|\partial_x\rho|^2
\end{equation}
for the active contribution to the interfacial tension (the isotropic active stress does not contribute). It takes on the same form that follows from the free energy Eq.~\eqref{eq:gl}. The coefficient $\kap_\text{A}=-2\al_dD_0v_0\tx^2\zeta<0$, however, is manifestly negative, and so is $\gam_\text{A}$. Using $\zeta\simeq v_0/(\mu_0\rho_0)$ for hard discs, we obtain $\kap_\text{A}=-2\al_d\kT(v_0\tx)^2/\rho_0$ which scales as $(v_0\tx)^2$ and thus dominates the contribution coming from the anisotropy of the stress $\msig_\text{IK}$ due to the interaction forces in agreement with the simulations~\cite{bial15}. For a simulation study including attractive forces see Ref.~\citenum{paliwal17}.

The force balance in the form of Eq.~\eqref{eq:j:p} offers another interpretation through treating $(v_0/\mu_0)\vec p$ as an external body force (like, \emph{e.g.}, gravity) caused by the polarization~\cite{steff17,epstein19,omar19}. The interfacial tension is then associated only with the anisotropy of the local interaction stresses. There are thus two interpretations yielding different expressions for the interfacial tension: either a non-vanishing polarization is maintained by internal stresses in the suspension or treated as an external body force.

\subsubsection{Coexistence and effective potentials}
\label{sec:coex}

Since $\nabla\cdot\msig=0$ holds for ABPs in the absence of particle currents, the component $\sig_\perp$ of the total stress along the normal has to be constant, $\sig_\perp(x)=-p^\ast$. Inspecting the explicit expression Eq.~\eqref{eq:sig:a}, we see that $\sig_\perp=-p(\rho)$ for homogeneous suspensions with (mechanical) pressure $p(\rho)$ depending only on the density [Eq.~\eqref{eq:pres}]. We thus obtain a first condition $p(\rho_+)=p(\rho_-)=p^\ast$ for the coexisting densities $\rho_\pm$, \emph{i.e.}, the bulk pressure has to be equal within each phase. However, this is not enough to unambiguously determine $\rho_\pm$. Performing a Maxwell construction fails for ABPs as demonstrated in computer simulations~\cite{solo15}.

To obtain a second condition, we follow Solon \emph{et al.}~\cite{solo17} and integrate $-\sig_\perp\partial_x\hat v$ with some function $\hat v(x)$,
\begin{equation}
  \label{eq:solon}
  \IInt{x}{-\infty}{+\infty} (p-\delta\sig_\perp)\partial_x\hat v = p^\ast(\hat v_+-\hat v_-),
\end{equation}
where $\delta\sig_\perp$ contains only derivatives and $\hat v_\pm$ indicates the value in the corresponding bulk phase. For the right hand side we have used $\sig_\perp(x)=-p^\ast$. The function $\hat v(x)$ is now chosen so that $\delta\sig_\perp\partial_x\hat v$ becomes a total derivative with respect to $x$ and, therefore, its integral vanishes. Writing the pressure $p=-\pd{(\hat v\phi)}{\hat v}$ as the conjugate observable to $\hat v$ with respect to $\phi(\hat v)$, Eq.~\eqref{eq:solon} yields the required second condition $\mu(\hat v_+)=\mu(\hat v_-)=\mu^\ast$ for the function $\mu(\hat v)=(\phi+p)\hat v=-\hat v^2\pd{\phi}{\hat v}$, which we recognize as an effective chemical potential.

The coexisting bulk densities of ABPs can thus be determined from an effective equilibrium system with free energy density $\phi(\hat v)$ but using an order parameter $\hat v=\hat v(\rho,\partial_x\rho,\dots)$ that plays the role of a local volume per particle. For this effective system we recover the conventional thermodynamic relations including the equality of pressure and chemical potential, giving rise to a Maxwell construction on $\phi(\hat v)$. The actual coexisting densities follow from inverting $\hat v_\pm=\hat v(\rho_\pm)$ and, in contrast to passive systems, now depend on properties of the interface.

The ``density'' $\hat\rho=1/\hat v$ minimizes the functional~\cite{solo17}
\begin{equation}
  \Phi[\hat\rho] = \Int{^d\x} \left\{\frac{\hat\kap(\hat\rho)}{2}|\nabla\hat\rho|^2 + \phi(\hat\rho)\right\}
  \label{eq:Phi}
\end{equation}
with $\fd{\Phi}{\hat\rho}=\mu^\ast$, replacing the free energy Eq.~\eqref{eq:gl}. Here, $\hat\kap>0$ is necessarily positive. Consequently, the interfacial tension $\hat\gam=\Int{x}\hat\kap|\partial_x\hat\rho|^2$ one would infer from this functional is also positive and different from $\gam_\text{A}$ obtained from the active stresses due to the polarization. This is not surprising since $\hat\gam$ is the tension corresponding to an effective equilibrium system governed by $\Phi$.

An alternative notion of a chemical potential for ABPs has been proposed in Ref.~\citenum{paliwal18} through identifying the particle current $j_x/\mu_0=\partial_x\sig_\perp(x)=-\rho\partial_x\mu(x)$ with the spatial derivative of a function $\mu(x)$ (clearly, for vanishing current $\mu=\mu^\ast$ has to be constant). An explicit expression for $\mu(x)$ can then be constructed from $\sig_\perp(x)$, which extends the conventional passive expression for the chemical potential by a swim potential~\cite{paliwal18}. Yet another approach along related lines but inspired by density functional theory has been proposed recently~\cite{herm19a}. It is based on splitting the conditional one-body force $\vec F=\vec F^\text{ad}+\vec F^\text{sup}$ [Eq.~\eqref{eq:F}] into intrinsic adiabatic forces $\vec F^\text{ad}$ and superadiabatic forces $\vec F^\text{sup}$. Also this approach constructs effective conservative forces that, in principle, yield the same density profile as the original ABPs. Again, the effective interfacial tension is positive and conceptually different from $\gam_\text{A}$~\cite{herm19b,wittmann19}.

To conclude, in contrast to passive suspensions, mechanical and (effective) thermodynamic route to the interfacial tension do not coincide for ABPs. This serves as a reminder that ABPs are driven away from thermal equilibrium even though certain aspects can be treated in analogy with equilibrium statistical mechanics.

\subsubsection{Scalar field theories}
\label{sec:field}

Alternatively, the large-scale behavior of ABPs can be approached adopting a top-down perspective in the spirit of field theories, which are based on symmetries and conservation laws while microscopic details are hidden in (unknown) coefficients. The starting point is the observation that, while the density is conserved, perturbations of the polarization [Eq.~\eqref{eq:p}] relax on the time scale $\tx$. This suggests a closure can be obtained by approximating the particle current $\vec j=\vec j(\rho,\nabla\rho,\dots)$ as a function of the local density $\rho(\x;t)$ and its derivatives. This ansatz can be made more rigorous through exploiting the large time-scale separation for the relaxation between density and polarization in the vicinity of the critical point~\cite{spec15}. To lowest order, one finds a current $\vec j=\vec j_\text{pot}=-\nabla\fd{\mathcal F}{\rho}$ with an effective potential of the form Eq.~\eqref{eq:gl} but with coefficients $\kap$, $a$, and $b$ that now depend on $v_0$ and the force imbalance coefficient $\zeta_\text{c}$ at the critical point. The large-scale behavior thus reduces to the passive ``Model B'' in the nomenclature of Hohenberg and Halperin~\cite{hohe77}, with the evolution of the density $\partial_t\rho=-\nabla\cdot\vec j$ given by a Cahn-Hilliard equation~\cite{spec14}. Consequently, the evolution of the (coarse-grained) density following from such an expansion obeys detailed balance while the microscopic equations~\eqref{eq:lang} do not.

\begin{figure}[t]
  \centering
  \includegraphics[width=\linewidth]{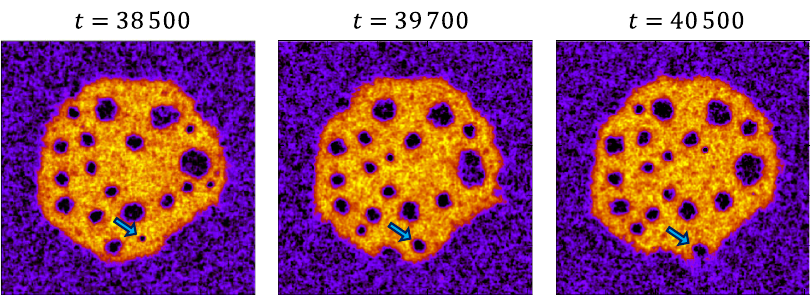}
  \caption{Active phase coexistence of dense liquid (yellow) and dilute gas phase (purple) as obtained from solving the evolution equation of Active Model B+ (with $\Lam=1$, $\xi=4$). The arrows mark the creation and expulsion of a bubble within the dense phase. Reproduced from Ref.~\citenum{tjhu18}.}
  \label{fig:bubbly}
\end{figure}

Cates and coworkers have explored the consequences of including higher-order derivatives $\vec j_\text{non}=-\Lam\nabla|\nabla\rho|^2+\xi(\nabla^2\rho)\nabla\rho$ to the current $\vec j=\vec j_\text{pot}+\vec j_\text{non}$ that cannot be cast into the functional derivative of a free energy. Hence, also the evolution of the density field now breaks detailed balance. For $\xi=0$ and $\Lam\neq 0$ (Active Model B), coexisting phases depend on $\Lam$ but can still be obtained from a modified common tangent construction~\cite{witt14}. With both terms contributing (Active Model B+), vapor bubbles are promoted (coined reverse Ostwald ripening), which lead to ``bubbly'' phase separation (Fig.~\ref{fig:bubbly}) and enable stable microphase separation of finite dense domains~\cite{tjhu18}. This mechanism can be captured through an effective interfacial tension that becomes negative. Pushing the systematic expansion of the hydrodynamic equations to higher orders reveals that many more terms contribute than those captured by $\Lam$ and $\zeta$ and, moreover, that they are not independent but related through the microscopic model parameters ($v_0$, $\tx$, $D_0$, \dots)~\cite{bickmann19,rapp19}.

\subsubsection{Non-uniform motility}

So far we have discussed inhomogeneous density profiles that emerge for spatially uniform motility, \emph{i.e.}, speed $v_0$ and correlation time $\tx$ do not depend on position. Another route that we briefly mention is to spatially modulate motility, which can be achieved experimentally for light-controlled Janus particles~\cite{pala13a,loza16}. In this case the pressure is non-uniform already for non-interacting ABPs~\cite{solo15a}. For discontinuous motility parameters one finds continuity conditions: for $D_0>0$ the density is continuous and from $v_0\vec p=\mu_0\nabla\cdot\msig_\text{A}$ we find that also the active stress is continuous across the boundary, $\vec n\cdot(\msig_\text{A}^>-\msig_\text{A}^<)=0$, with boundary normal $\vec n$. The later places a continuity condition on the derivative of the polarization, which thus jumps across the boundary. We again find a non-zero polarization confined to a finite interfacial region separating the two regions with different bulk densities~\cite{fisc20}.

\subsection{Forces on immersed objects}
\label{sec:imm}

\begin{figure*}[t]
  \centering
  \includegraphics{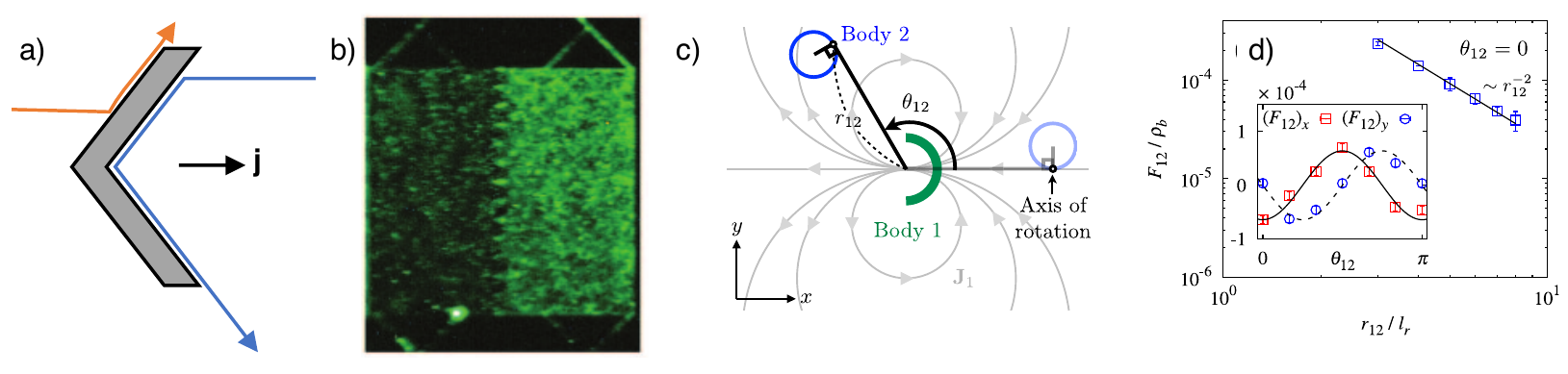}
  \caption{Asymmetric obstacles. (a)~Sketch of two trajectories (with large persistence length) coming from both sides of a chevron. It is easy to see that blue trajectories are trapped a longer time (on average), and that the obstacle generates a net current $\vec j$. (b)~Snapshot of an experiment with bacteria, where a row of fixed chevrons has been placed along the center of the compartment. The current through the chevrons leads to a steady state with two different densities. Reproduced from Ref.~\citenum{galajda07}. (c)~Sketch of the currents generated by a ``boomerang'' at the center. (d)~These currents mediate a force $F_{12}$ on a disc that decays as a power law $~r_{12}^{-2}$ in two dimensions with distance $r$ and depends on the angle $\theta_{12}$ with the symmetry axis (inset). Reproduced from Ref.~\citenum{baek18}.}
  \label{fig:chevron}
\end{figure*}

An object immersed into a passive liquid or suspension (so that it is completely surrounded) will not experience a net force irrespective of its shape. This might be different for active suspensions, in which asymmetric shapes can generate persistent steady currents of active particles. An example is shown in Fig.~\ref{fig:chevron}a for a chevron. Active particles coming from the right are trapped for some time due to their persistence of motion~\cite{kais12} while particles coming from the left are deflected. This implies an effective particle current $\vec j$ to the right. Placing a number of fixed obstacles with such a shape can be used to generate a density gradient~\cite{galajda07,wan08,sten16}, cf. Fig.~\ref{fig:chevron}b. Another example is shown in Fig.~\ref{fig:chevron}c for a ``boomerang'' showing the induced currents of active particles.

Isolated bodies that do not generate currents are still force-free in suspensions of ABPs. This can be seen immediately from the force balance Eq.~\eqref{eq:j:sig} with $\vec j=0$, for which $\vec F^\text{body}=-\IInt{^d\x}{V}{}\vec F^\text{ex}\rho=\oint_{\partial V}\text{d}^{d-1}\x\;\vec n\cdot\msig$ for any integration volume $V$ that fully encloses the body. Expanding the volume so that its surface $\partial V$ completely lies within the uniform suspension ($\vec p=0$ and $\vec Q=0$) with diagonal $\msig$ [cf. Eq.~\eqref{eq:sig:a}] yields $\vec F^\text{body}=0$. Of course, this might not be possible in the presence of walls or other bodies, leading to depletion-like interactions. These have been studied in computer simulations for bodies with different shapes such as parallel walls~\cite{ni15}, discs~\cite{yamchi17}, and rods~\cite{harder14}.

Conversely, if a shape generates currents then these currents exert a body force (in an infinite system) $\vec F^\text{body}=-\IInt{^d\x}{V}{}\vec j/\mu_0$~\cite{niko16}. For non-interacting ABPs in $d=2$ dimensions, the force generated through an asymmetric body onto a second body has been calculated in Ref.~\cite{baek18}. This force depends on the angle with the body's symmetry axis and decays as a power law $\sim r^{-2}$ with the distance $r$ between the bodies [Fig.~\ref{fig:chevron}d]. Since the force back onto the first body now depends on the shape of the second body, Newton's third law can be broken for the effective interactions (on the level of the immersed bodies), which has interesting general consequences~\cite{ivlev15}. Estimating the excess density on the surface of bodies allows to directly calculate $\vec F^\text{body}$~\cite{yan15a}. Beyond rigid bodies there are first numerical studies on the dynamics of immersed flexible filaments~\cite{niko16,shin17} and membranes~\cite{mallory15}. Finally, we note that beyond the forces sustained by currents there can also be collective forces due to fluctuations that are enhanced by the activity~\cite{ray14,rohwer17}.


\section{Thermodynamics}

\subsection{Time-reversal symmetry}

Since active matter can generate non-trivial collective forces on walls and immersed objects, it seems natural to exploit these forces in order to perform useful work. This leads to questions like how much work is available, what is the efficiency, and, more generally, how to quantify entropy production and dissipation in active suspensions.

While no thermodynamical potential (in the conventional sense) exists for steadily driven systems, notions like work, heat, and entropy production are still meaningful. Stochastic thermodynamics is a powerful framework to study these notions in driven systems that are dominated by fluctuations~\cite{seif12}. Driving a system through time-dependent changes of parameters and non-conservative forces, it can be shown that the dissipated heat $q$ is related to the breaking of time-reversal symmetry,
\begin{equation}
  \frac{q}{\kT} = \Delta s_\text{m} = -\ln\frac{\mathcal P[X]}{\mathcal P[X^\dagger]},
  \label{eq:trs}
\end{equation}
where $T$ is the temperature of the environment into which the heat is dissipated and $\Delta s_\text{m}$ is the corresponding increase of (dimensionless) entropy in the environment. Eq.~\eqref{eq:trs} holds for single stochastic trajectories $X=\{x_t\}_0^\tau$ of length $\tau$ with path probability $\mathcal P[X]$. This is to be compared to the probability of observing the time-reversed trajectory $X^\dagger=\{x^\dagger_{\tau-t}\}_0^\tau$. It is important to note that for the equalities in Eq.~\eqref{eq:trs} to hold $x$ needs to encompass all degrees of freedom that contribute to the entropy production~\cite{seif19}.

Nevertheless, Eq.~\eqref{eq:trs} has sparked quite some interest in the context of active particles and has been applied to Eq.~\eqref{eq:lang} yielding some measure $\Delta\tilde s_\text{m}$ for the breaking of time-reversal symmetry. One strategy has been to map a variant of ABPs called active Ornstein-Uhlenbeck process (AOUP) to second-order stochastic differential equations and to interpret those within stochastic thermodynamics~\cite{fodo16,mand17}, see also the the comment Ref.~\citenum{caprini18}. Moreover, links to information-theoretical arguments have been drawn~\cite{dabe19}. How active particles break time-reversal symmetry has also been extended to the field theories sketched in Sec.~\ref{sec:field}~\cite{nard17}.

The major conceptual difficulty is that the behavior of the particle orientation under time reversal is not fixed by the model but has to be supplied: either odd ($\vec e_i^\dagger=-\vec e_i$) corresponding to interpreting $v_0\vec e_i$ as a solvent velocity or even ($\vec e_i^\dagger=\vec e_i$) with $(v_0/\mu_0)\vec e_i$ behaving as a non-conservative (stochastic) force~\cite{gang13,fala16,spec16}. Clearly, this choice leads to different expressions for $\Delta\tilde s_\text{m}$~\cite{marc17,puglisi17,shankar18,crosato19}.

\subsection{Entropy production and dissipation}

To be distinguished from breaking time-reversal symmetry on the level of particle trajectories is the actual entropy production and dissipated heat $q$ required to maintain a suspension of active particles away from equilibrium. Thermodynamic consistency requires that directed motion fueled by some (free) energy (typically a difference $\Delta\mu$ of chemical potential) couples to the potential energy $U$~\footnote{Note that $\Delta\mu$ relates to the chemical potential difference of two chemical species and is not to be confused with the effective chemical potential in Sec.~\ref{sec:coex}.}. In particular, if a particle has to go against a force then it slows down for a fixed energy budget $\Delta\mu$. Interestingly, this also implies that dragging an active particle can synthesize fuel molecules~\cite{gasp17}.

To proceed, we need to take into account the actual propulsion mechanism. On a schematic level, the simplest model is discrete jumps of length $\lam$ with rate $\kap^+_i$ along the orientation $\vec e_i$ and $\kap^-_i$ against the orientation. Every jump corresponds to a chemical event that consumes a fuel molecule through translating particle $i$. The rates obey the local detailed balance condition, $\kap^+_i/\kap^-_i=e^{(\Delta\mu-\lam\vec e_i\cdot\nabla_i U)/\kT}$, where we have used that $\lam$ is small compared to other length scales (in particular the diameter of active particles)~\cite{spec18}. This coupling implies that externally forcing a particle allows to revers the reaction and to synthesize fuel molecules~\cite{gasp17}. Pietzonka and Seifert have derived and analyzed the continuum equations of a lattice model with lattice spacing $\lam$ in the limit $\lam\to0$~\cite{piet17}. A different approach is to eliminate the chemical degrees of freedom for finite but small $\lam$, which yields a simple variation of Eq.~\eqref{eq:lang} in which the speed $v_0$ is replaced by an expression that explicitly depends on the change of potential energy~\cite{spec18}.

If $\dot n=\sum_i\dot n_i$ denotes the total number of chemical events per time then  the work per time spent on driving the particles is $\dot w=\dot n\Delta\mu$. Assuming symmetric rates with attempt rate $\kap_0$, for particle $i$ to lowest order the average contribution to the work reads~\cite{spec18}
\begin{equation}
  \mean{\dot w_i} \approx \mean{v_i\vec e_i\cdot\nabla_i U} + \kap_0\mean{(\Delta\mu-\lam\vec e_i\cdot\nabla_i U)^2}/\kT,
\end{equation}
where $v_i=\lam\dot n_i$ is the actual propulsion speed. The first term is what is expected for the work from breaking time-reversal symmetry if treating $v_i\vec e_i$ as an odd speed, $\mean{\dot w_i^\text{ABP}}=\mean{v_0\vec e_i\cdot\nabla_iU}$, which thus fixes the prescription for ABPs and questions the consistent interpretation of the polarization as a body force in Eq.~\eqref{eq:j:sig}. This work is a lower bound to the actual work, $\mean{\dot w_i}\geqslant\mean{\dot w_i^\text{ABP}}$, the leading correction of which takes the form of a variance of energy fluctuations around $\Delta\mu$ (reminiscent of energy fluctuations in the isobaric ensemble). Some of the dissipated work can be extracted again due to the forces on immersed bodies outlined in Sec.~\ref{sec:imm}~\cite{piet19}.


\section{Perspectives}

Collective forces are involved in the wealth of dynamic collective behavior that is observed in active matter. There are two main thrusts: First, the understanding of these forces is pivotal for modeling the effective coarse-grained dynamics of these systems, which is independent of many microscopic details. Second, the directed motion of active constituents (synthetic particles, bacteria, molecular motors) exerts collective forces on the environment that can be harvested as useful work, \emph{e.g.} through ``engines'' powered by active particles~\cite{leon10,krishna16,vizs17} or to enable novel material responses to external perturbations (such as active metamaterials)~\cite{souslov17}.

The theoretical ideas we have reviewed here will be useful to design and optimize interactions of active systems with their environment. One application is the autonomous self-assembly of ordered target structures from disordered molecular or colloidal building blocks. For dynamics obeying detailed balance, self-assembly is purely driven by the gradient of free energy, leaving a small window within which the competition of non-specific and specific interactions allow successful assembly~\cite{whit09}. Doting with active particles induces stresses that have been shown to speed up colloidal crystallization and to potentially broaden this window~\cite{kumm15,meer16,mallory19}. Moreover, it might be possible to realize (metastable) ordered structures that are inaccessible with dynamics obeying detailed balance.

Self-assembly of biological matter yields functioning molecular ``structures'' such as enzymes which, in addition, are typically hierarchical. These can be characterized as ``machines of machines'' in contrast to artificial engines built from inert parts~\cite{needleman17}. Simplified synthetic systems that mimic such processes can yield valuable insights into the underlying physical principles. One route is the assembly of active but individually immotile components into small clusters that exhibit translation and rotation, the fundamental forms of motility. Specific examples include metallic rods~\cite{wykes16} and spherical ion-exchange particles~\cite{niu18}. The next step will be to synthesize engines that exert forces and thus manipulate their environment. Coupled engines, each with a limited set of responses, that form patterns which are switchable could perform complex tasks like sorting and cargo delivery as demonstrated by robots~\cite{xie19,li19}.

Membrane-less organelles -- inhomogeneous concentrations of proteins and polymers -- have turned out to be a major building block to organize and compartmentalize biological reactions within the cell and the nucleus~\cite{bran15}. These organelles have been recognized as liquid droplets stabilized by an interfacial tension~\cite{bran09}. Quite remarkable, the cell thus exploits a basic physical mechanism to acquire and release reactants and product molecules. Factors like viscoelasticity and intrinsic non-equilibrium features of the cytoplasm (\emph{e.g.}, motor proteins organizing the cytoskeleton), however, will require to adapt and extend the existing theoretical framework for phase separation of passive suspensions and mixtures~\cite{hyman14}. Developing a comprehensive understanding of the non-equilibrium coexistence in ABPs is one step in this direction. Going beyond single cells, aggregates of cells organizing into, \emph{e.g.}, tissues again exert forces which can be measured experimentally through traction force microscopy~\cite{sabass08,trepat09}.


\section{Conclusions}

We have reviewed the forces exerted by a steady-state suspension of active Brownian particles on walls and immersed bodies, and stabilizing density inhomogeneities. These forces are caused by a non-vanishing polarization, which, for ABPs, is not caused by alignment but trapping, and is accompanied by density gradients. The polarization can then be expressed as an active stress, Eq.~\eqref{eq:sig:a}. ABPs clearly break detailed balance and thus time-reversal symmetry. We have clarified the distinction between breaking time-reversal symmetry on the level of (observable) particle trajectories and the dissipation required to maintain the suspension away from equilibrium. For the model of ABPs reviewed here, a comprehensive and unifying picture of collective dynamics, forces, and thermodynamics is now coming together. This model thus provides a reference to develop the formalism for more complex systems in which alignment, hydrodynamic coupling, and long-ranged phoretic forces can no longer be neglected.

\section*{Conflicts of interest}

There are no conflicts to declare.

\section*{Acknowledgements}

I thank H. Löwen, C. Bechinger, R. Winkler, R.L. Jack, C. Patrick Royall, T. Palberg, U. Seifert, F. Schmid, and P. Virnau for many inspiring discussions over the years. This manuscript would not have been possible without the countless discussions with students working on related projects: J. Bialk\'e, M. Rein, J. Siebers, A. Fischer, A. Jayaram, and M. Campo. Financial support is acknowledged by the Deutsche Forschungsgemeinschaft through the priority program SPP 1726 (grant no. 254473714).


\providecommand*{\mcitethebibliography}{\thebibliography}
\csname @ifundefined\endcsname{endmcitethebibliography}
{\let\endmcitethebibliography\endthebibliography}{}


\begin{mcitethebibliography}{132}
\providecommand*{\natexlab}[1]{#1}
\providecommand*{\mciteSetBstSublistMode}[1]{}
\providecommand*{\mciteSetBstMaxWidthForm}[2]{}
\providecommand*{\mciteBstWouldAddEndPuncttrue}
  {\def\EndOfBibitem{\unskip.}}
\providecommand*{\mciteBstWouldAddEndPunctfalse}
  {\let\EndOfBibitem\relax}
\providecommand*{\mciteSetBstMidEndSepPunct}[3]{}
\providecommand*{\mciteSetBstSublistLabelBeginEnd}[3]{}
\providecommand*{\EndOfBibitem}{}
\mciteSetBstSublistMode{f}
\mciteSetBstMaxWidthForm{subitem}
{(\emph{\alph{mcitesubitemcount}})}
\mciteSetBstSublistLabelBeginEnd{\mcitemaxwidthsubitemform\space}
{\relax}{\relax}

\bibitem[Gompper \emph{et~al.}(2019)Gompper, Winkler, Speck, Solon, Nardini,
  Peruani, Loewen, Golestanian, Kaupp, Alvarez, Kioerboe, Lauga, Poon, Simone,
  Cichos, Fischer, Landin, Soeker, Kapral, Gaspard, Ripoll, Sagues, Yeomans,
  Doostmohammadi, Aronson, Bechinger, Stark, Hemelrijk, Nedelec, Sarkar,
  Aryaksama, Lacroix, Duclos, Yashunsky, Silberzan, Arroyo, and Kale]{roadmap}
G.~Gompper, R.~G. Winkler, T.~Speck, A.~Solon, C.~Nardini, F.~Peruani,
  H.~Loewen, R.~Golestanian, U.~B. Kaupp, L.~Alvarez, T.~Kioerboe, E.~Lauga,
  W.~Poon, A.~D. Simone, F.~Cichos, A.~Fischer, S.~M. Landin, N.~Soeker,
  R.~Kapral, P.~Gaspard, M.~Ripoll, F.~Sagues, J.~Yeomans, A.~Doostmohammadi,
  I.~Aronson, C.~Bechinger, H.~Stark, C.~Hemelrijk, F.~Nedelec, T.~Sarkar,
  T.~Aryaksama, M.~Lacroix, G.~Duclos, V.~Yashunsky, P.~Silberzan, M.~Arroyo
  and S.~Kale, \emph{arXiv:1912.06710}, 2019\relax
\mciteBstWouldAddEndPuncttrue
\mciteSetBstMidEndSepPunct{\mcitedefaultmidpunct}
{\mcitedefaultendpunct}{\mcitedefaultseppunct}\relax
\EndOfBibitem
\bibitem[Basu \emph{et~al.}(2015)Basu, Maes, and Neto\ifmmode~\check{c}\else
  \v{c}\fi{}n\'y]{basu15a}
U.~Basu, C.~Maes and K.~Neto\ifmmode~\check{c}\else \v{c}\fi{}n\'y, \emph{Phys.
  Rev. Lett.}, 2015, \textbf{114}, 250601\relax
\mciteBstWouldAddEndPuncttrue
\mciteSetBstMidEndSepPunct{\mcitedefaultmidpunct}
{\mcitedefaultendpunct}{\mcitedefaultseppunct}\relax
\EndOfBibitem
\bibitem[Kardar and Golestanian(1999)]{kardar99}
M.~Kardar and R.~Golestanian, \emph{Rev. Mod. Phys.}, 1999, \textbf{71},
  1233--1245\relax
\mciteBstWouldAddEndPuncttrue
\mciteSetBstMidEndSepPunct{\mcitedefaultmidpunct}
{\mcitedefaultendpunct}{\mcitedefaultseppunct}\relax
\EndOfBibitem
\bibitem[Hertlein \emph{et~al.}(2008)Hertlein, Helden, Gambassi, Dietrich, and
  Bechinger]{hert08}
C.~Hertlein, L.~Helden, A.~Gambassi, S.~Dietrich and C.~Bechinger,
  \emph{Nature}, 2008, \textbf{451}, 172--175\relax
\mciteBstWouldAddEndPuncttrue
\mciteSetBstMidEndSepPunct{\mcitedefaultmidpunct}
{\mcitedefaultendpunct}{\mcitedefaultseppunct}\relax
\EndOfBibitem
\bibitem[Wu \emph{et~al.}(2017)Wu, Hishamunda, Chen, DeCamp, Chang,
  Fern{\'{a}}ndez-Nieves, Fraden, and Dogic]{wu17}
K.-T. Wu, J.~B. Hishamunda, D.~T.~N. Chen, S.~J. DeCamp, Y.-W. Chang,
  A.~Fern{\'{a}}ndez-Nieves, S.~Fraden and Z.~Dogic, \emph{Science}, 2017,
  \textbf{355}, eaal1979\relax
\mciteBstWouldAddEndPuncttrue
\mciteSetBstMidEndSepPunct{\mcitedefaultmidpunct}
{\mcitedefaultendpunct}{\mcitedefaultseppunct}\relax
\EndOfBibitem
\bibitem[Di~Leonardo \emph{et~al.}(2010)Di~Leonardo, Angelani, Dell'Arciprete,
  Ruocco, Iebba, Schippa, Conte, Mecarini, De~Angelis, and Di~Fabrizio]{leon10}
R.~Di~Leonardo, L.~Angelani, D.~Dell'Arciprete, G.~Ruocco, V.~Iebba,
  S.~Schippa, M.~P. Conte, F.~Mecarini, F.~De~Angelis and E.~Di~Fabrizio,
  \emph{Proc. Natl. Acad. Sci. U.S.A.}, 2010, \textbf{107}, 9541--9545\relax
\mciteBstWouldAddEndPuncttrue
\mciteSetBstMidEndSepPunct{\mcitedefaultmidpunct}
{\mcitedefaultendpunct}{\mcitedefaultseppunct}\relax
\EndOfBibitem
\bibitem[Sokolov \emph{et~al.}(2010)Sokolov, Apodaca, Grzybowski, and
  Aranson]{sokolov10}
A.~Sokolov, M.~M. Apodaca, B.~A. Grzybowski and I.~S. Aranson, \emph{Proc.
  Natl. Acad. Sci. U.S.A.}, 2010, \textbf{107}, 969--974\relax
\mciteBstWouldAddEndPuncttrue
\mciteSetBstMidEndSepPunct{\mcitedefaultmidpunct}
{\mcitedefaultendpunct}{\mcitedefaultseppunct}\relax
\EndOfBibitem
\bibitem[Aubret \emph{et~al.}(2018)Aubret, Youssef, Sacanna, and
  Palacci]{aubret18}
A.~Aubret, M.~Youssef, S.~Sacanna and J.~Palacci, \emph{Nat. Phys.}, 2018,
  \textbf{14}, 1114--1118\relax
\mciteBstWouldAddEndPuncttrue
\mciteSetBstMidEndSepPunct{\mcitedefaultmidpunct}
{\mcitedefaultendpunct}{\mcitedefaultseppunct}\relax
\EndOfBibitem
\bibitem[Buttinoni \emph{et~al.}(2013)Buttinoni, Bialk\'e, K\"ummel, L\"owen,
  Bechinger, and Speck]{butt13}
I.~Buttinoni, J.~Bialk\'e, F.~K\"ummel, H.~L\"owen, C.~Bechinger and T.~Speck,
  \emph{Phys. Rev. Lett.}, 2013, \textbf{110}, 238301\relax
\mciteBstWouldAddEndPuncttrue
\mciteSetBstMidEndSepPunct{\mcitedefaultmidpunct}
{\mcitedefaultendpunct}{\mcitedefaultseppunct}\relax
\EndOfBibitem
\bibitem[Chat\'{e} and Mahault(2019)]{chate19}
H.~Chat\'{e} and B.~Mahault, \emph{arXiv:1906.05542}, 2019\relax
\mciteBstWouldAddEndPuncttrue
\mciteSetBstMidEndSepPunct{\mcitedefaultmidpunct}
{\mcitedefaultendpunct}{\mcitedefaultseppunct}\relax
\EndOfBibitem
\bibitem[Lauga and Powers(2009)]{laug09}
E.~Lauga and T.~R. Powers, \emph{Rep. Prog. Phys.}, 2009, \textbf{72},
  096601\relax
\mciteBstWouldAddEndPuncttrue
\mciteSetBstMidEndSepPunct{\mcitedefaultmidpunct}
{\mcitedefaultendpunct}{\mcitedefaultseppunct}\relax
\EndOfBibitem
\bibitem[Elgeti \emph{et~al.}(2015)Elgeti, Winkler, and Gompper]{elge15}
J.~Elgeti, R.~G. Winkler and G.~Gompper, \emph{Rep. Prog. Phys.}, 2015,
  \textbf{78}, 056601\relax
\mciteBstWouldAddEndPuncttrue
\mciteSetBstMidEndSepPunct{\mcitedefaultmidpunct}
{\mcitedefaultendpunct}{\mcitedefaultseppunct}\relax
\EndOfBibitem
\bibitem[Bechinger \emph{et~al.}(2016)Bechinger, Leonardo, L\"owen, Reichhardt,
  Volpe, and Volpe]{bech16}
C.~Bechinger, R.~D. Leonardo, H.~L\"owen, C.~Reichhardt, G.~Volpe and G.~Volpe,
  \emph{Rev. Mod. Phys.}, 2016, \textbf{88}, 045006\relax
\mciteBstWouldAddEndPuncttrue
\mciteSetBstMidEndSepPunct{\mcitedefaultmidpunct}
{\mcitedefaultendpunct}{\mcitedefaultseppunct}\relax
\EndOfBibitem
\bibitem[Zöttl and Stark(2016)]{zott16}
A.~Zöttl and H.~Stark, \emph{J. Phys. Condens. Matter}, 2016, \textbf{28},
  253001\relax
\mciteBstWouldAddEndPuncttrue
\mciteSetBstMidEndSepPunct{\mcitedefaultmidpunct}
{\mcitedefaultendpunct}{\mcitedefaultseppunct}\relax
\EndOfBibitem
\bibitem[Cates and Tailleur(2015)]{cate15}
M.~E. Cates and J.~Tailleur, \emph{Annu. Rev. Condens. Matter Phys.}, 2015,
  \textbf{6}, 219--244\relax
\mciteBstWouldAddEndPuncttrue
\mciteSetBstMidEndSepPunct{\mcitedefaultmidpunct}
{\mcitedefaultendpunct}{\mcitedefaultseppunct}\relax
\EndOfBibitem
\bibitem[Prost \emph{et~al.}(2015)Prost, J\"ulicher, and Joanny]{pros15}
J.~Prost, F.~J\"ulicher and J.-F. Joanny, \emph{Nat. Phys.}, 2015, \textbf{11},
  111--117\relax
\mciteBstWouldAddEndPuncttrue
\mciteSetBstMidEndSepPunct{\mcitedefaultmidpunct}
{\mcitedefaultendpunct}{\mcitedefaultseppunct}\relax
\EndOfBibitem
\bibitem[Needleman and Dogic(2017)]{needleman17}
D.~Needleman and Z.~Dogic, \emph{Nat. Rev. Mater.}, 2017, \textbf{2},
  17048\relax
\mciteBstWouldAddEndPuncttrue
\mciteSetBstMidEndSepPunct{\mcitedefaultmidpunct}
{\mcitedefaultendpunct}{\mcitedefaultseppunct}\relax
\EndOfBibitem
\bibitem[Kubo(1966)]{kubo66}
R.~Kubo, \emph{Rep. Prog. Phys.}, 1966, \textbf{29}, 255\relax
\mciteBstWouldAddEndPuncttrue
\mciteSetBstMidEndSepPunct{\mcitedefaultmidpunct}
{\mcitedefaultendpunct}{\mcitedefaultseppunct}\relax
\EndOfBibitem
\bibitem[Bruss and Glotzer(2018)]{bruss18}
I.~R. Bruss and S.~C. Glotzer, \emph{Phys. Rev. E}, 2018, \textbf{97},
  042609\relax
\mciteBstWouldAddEndPuncttrue
\mciteSetBstMidEndSepPunct{\mcitedefaultmidpunct}
{\mcitedefaultendpunct}{\mcitedefaultseppunct}\relax
\EndOfBibitem
\bibitem[Reichhardt and Reichhardt(2018)]{reich18}
C.~Reichhardt and C.~J.~O. Reichhardt, \emph{Phys. Rev. E}, 2018, \textbf{97},
  052613\relax
\mciteBstWouldAddEndPuncttrue
\mciteSetBstMidEndSepPunct{\mcitedefaultmidpunct}
{\mcitedefaultendpunct}{\mcitedefaultseppunct}\relax
\EndOfBibitem
\bibitem[Howse \emph{et~al.}(2007)Howse, Jones, Ryan, Gough, Vafabakhsh, and
  Golestanian]{hows07}
J.~R. Howse, R.~A.~L. Jones, A.~J. Ryan, T.~Gough, R.~Vafabakhsh and
  R.~Golestanian, \emph{Phys. Rev. Lett.}, 2007, \textbf{99}, 048102\relax
\mciteBstWouldAddEndPuncttrue
\mciteSetBstMidEndSepPunct{\mcitedefaultmidpunct}
{\mcitedefaultendpunct}{\mcitedefaultseppunct}\relax
\EndOfBibitem
\bibitem[Wagner \emph{et~al.}(2017)Wagner, Hagan, and Baskaran]{wagner17}
C.~G. Wagner, M.~F. Hagan and A.~Baskaran, \emph{J. Stat. Mech.: Theory Exp.},
  2017, \textbf{2017}, 043203\relax
\mciteBstWouldAddEndPuncttrue
\mciteSetBstMidEndSepPunct{\mcitedefaultmidpunct}
{\mcitedefaultendpunct}{\mcitedefaultseppunct}\relax
\EndOfBibitem
\bibitem[Hermann and Schmidt(2018)]{hermann18}
S.~Hermann and M.~Schmidt, \emph{Soft Matter}, 2018, \textbf{14},
  1614--1621\relax
\mciteBstWouldAddEndPuncttrue
\mciteSetBstMidEndSepPunct{\mcitedefaultmidpunct}
{\mcitedefaultendpunct}{\mcitedefaultseppunct}\relax
\EndOfBibitem
\bibitem[Basu \emph{et~al.}(2018)Basu, Majumdar, Rosso, and Schehr]{basu18}
U.~Basu, S.~N. Majumdar, A.~Rosso and G.~Schehr, \emph{Phys. Rev. E}, 2018,
  \textbf{98}, 062121\relax
\mciteBstWouldAddEndPuncttrue
\mciteSetBstMidEndSepPunct{\mcitedefaultmidpunct}
{\mcitedefaultendpunct}{\mcitedefaultseppunct}\relax
\EndOfBibitem
\bibitem[Basu \emph{et~al.}(2019)Basu, Majumdar, Rosso, and Schehr]{basu19}
U.~Basu, S.~N. Majumdar, A.~Rosso and G.~Schehr, \emph{Phys. Rev. E}, 2019,
  \textbf{100}, 062116\relax
\mciteBstWouldAddEndPuncttrue
\mciteSetBstMidEndSepPunct{\mcitedefaultmidpunct}
{\mcitedefaultendpunct}{\mcitedefaultseppunct}\relax
\EndOfBibitem
\bibitem[Irving and Kirkwood(1950)]{irving50}
J.~H. Irving and J.~G. Kirkwood, \emph{J. Chem. Phys.}, 1950, \textbf{18},
  817--829\relax
\mciteBstWouldAddEndPuncttrue
\mciteSetBstMidEndSepPunct{\mcitedefaultmidpunct}
{\mcitedefaultendpunct}{\mcitedefaultseppunct}\relax
\EndOfBibitem
\bibitem[Saintillan and Shelley(2015)]{saint15}
D.~Saintillan and M.~J. Shelley, \emph{Complex Fluids in Biological Systems},
  Springer, New York, NY, 2015, pp. 319--355\relax
\mciteBstWouldAddEndPuncttrue
\mciteSetBstMidEndSepPunct{\mcitedefaultmidpunct}
{\mcitedefaultendpunct}{\mcitedefaultseppunct}\relax
\EndOfBibitem
\bibitem[Bialk\'e \emph{et~al.}(2013)Bialk\'e, L\"owen, and Speck]{bial13}
J.~Bialk\'e, H.~L\"owen and T.~Speck, \emph{EPL}, 2013, \textbf{103},
  30008\relax
\mciteBstWouldAddEndPuncttrue
\mciteSetBstMidEndSepPunct{\mcitedefaultmidpunct}
{\mcitedefaultendpunct}{\mcitedefaultseppunct}\relax
\EndOfBibitem
\bibitem[Speck \emph{et~al.}(2015)Speck, Menzel, Bialk\'e, and L\"owen]{spec15}
T.~Speck, A.~M. Menzel, J.~Bialk\'e and H.~L\"owen, \emph{J. Chem. Phys.},
  2015, \textbf{142}, 224109\relax
\mciteBstWouldAddEndPuncttrue
\mciteSetBstMidEndSepPunct{\mcitedefaultmidpunct}
{\mcitedefaultendpunct}{\mcitedefaultseppunct}\relax
\EndOfBibitem
\bibitem[Stenhammar \emph{et~al.}(2013)Stenhammar, Tiribocchi, Allen,
  Marenduzzo, and Cates]{sten13}
J.~Stenhammar, A.~Tiribocchi, R.~J. Allen, D.~Marenduzzo and M.~E. Cates,
  \emph{Phys. Rev. Lett.}, 2013, \textbf{111}, 145702\relax
\mciteBstWouldAddEndPuncttrue
\mciteSetBstMidEndSepPunct{\mcitedefaultmidpunct}
{\mcitedefaultendpunct}{\mcitedefaultseppunct}\relax
\EndOfBibitem
\bibitem[H\"artel \emph{et~al.}(2018)H\"artel, Richard, and Speck]{hart18}
A.~H\"artel, D.~Richard and T.~Speck, \emph{Phys. Rev. E}, 2018, \textbf{97},
  012606\relax
\mciteBstWouldAddEndPuncttrue
\mciteSetBstMidEndSepPunct{\mcitedefaultmidpunct}
{\mcitedefaultendpunct}{\mcitedefaultseppunct}\relax
\EndOfBibitem
\bibitem[Arnoulx~de Pirey \emph{et~al.}(2019)Arnoulx~de Pirey, Lozano, and van
  Wijland]{pirey19}
T.~Arnoulx~de Pirey, G.~Lozano and F.~van Wijland, \emph{Phys. Rev. Lett.},
  2019, \textbf{123}, 260602\relax
\mciteBstWouldAddEndPuncttrue
\mciteSetBstMidEndSepPunct{\mcitedefaultmidpunct}
{\mcitedefaultendpunct}{\mcitedefaultseppunct}\relax
\EndOfBibitem
\bibitem[Bertin \emph{et~al.}(2006)Bertin, Droz, and Gr\'egoire]{bert06}
E.~Bertin, M.~Droz and G.~Gr\'egoire, \emph{Phys. Rev. E}, 2006, \textbf{74},
  022101\relax
\mciteBstWouldAddEndPuncttrue
\mciteSetBstMidEndSepPunct{\mcitedefaultmidpunct}
{\mcitedefaultendpunct}{\mcitedefaultseppunct}\relax
\EndOfBibitem
\bibitem[Epstein \emph{et~al.}(2019)Epstein, Klymko, and Mandadapu]{epstein19}
J.~M. Epstein, K.~Klymko and K.~K. Mandadapu, \emph{The Journal of Chemical
  Physics}, 2019, \textbf{150}, 164111\relax
\mciteBstWouldAddEndPuncttrue
\mciteSetBstMidEndSepPunct{\mcitedefaultmidpunct}
{\mcitedefaultendpunct}{\mcitedefaultseppunct}\relax
\EndOfBibitem
\bibitem[Steffenoni \emph{et~al.}(2017)Steffenoni, Falasco, and Kroy]{steff17}
S.~Steffenoni, G.~Falasco and K.~Kroy, \emph{Phys. Rev. E}, 2017, \textbf{95},
  052142\relax
\mciteBstWouldAddEndPuncttrue
\mciteSetBstMidEndSepPunct{\mcitedefaultmidpunct}
{\mcitedefaultendpunct}{\mcitedefaultseppunct}\relax
\EndOfBibitem
\bibitem[Farage \emph{et~al.}(2015)Farage, Krinninger, and Brader]{fara15}
T.~F.~F. Farage, P.~Krinninger and J.~M. Brader, \emph{Phys. Rev. E}, 2015,
  \textbf{91}, 042310\relax
\mciteBstWouldAddEndPuncttrue
\mciteSetBstMidEndSepPunct{\mcitedefaultmidpunct}
{\mcitedefaultendpunct}{\mcitedefaultseppunct}\relax
\EndOfBibitem
\bibitem[Maggi \emph{et~al.}(2015)Maggi, Marconi, Gnan, and
  Di~Leonardo]{magg15}
C.~Maggi, U.~M.~B. Marconi, N.~Gnan and R.~Di~Leonardo, \emph{Sci. Rep.}, 2015,
  \textbf{5}, 10742\relax
\mciteBstWouldAddEndPuncttrue
\mciteSetBstMidEndSepPunct{\mcitedefaultmidpunct}
{\mcitedefaultendpunct}{\mcitedefaultseppunct}\relax
\EndOfBibitem
\bibitem[Marconi and Maggi(2015)]{marc15}
U.~M.~B. Marconi and C.~Maggi, \emph{Soft Matter}, 2015, \textbf{11},
  8768--8781\relax
\mciteBstWouldAddEndPuncttrue
\mciteSetBstMidEndSepPunct{\mcitedefaultmidpunct}
{\mcitedefaultendpunct}{\mcitedefaultseppunct}\relax
\EndOfBibitem
\bibitem[Rein and Speck(2016)]{rein16a}
M.~Rein and T.~Speck, \emph{Eur. Phys. J. E}, 2016, \textbf{39}, 84\relax
\mciteBstWouldAddEndPuncttrue
\mciteSetBstMidEndSepPunct{\mcitedefaultmidpunct}
{\mcitedefaultendpunct}{\mcitedefaultseppunct}\relax
\EndOfBibitem
\bibitem[Speck and Jack(2016)]{spec16a}
T.~Speck and R.~L. Jack, \emph{Phys. Rev. E}, 2016, \textbf{93}, 062605\relax
\mciteBstWouldAddEndPuncttrue
\mciteSetBstMidEndSepPunct{\mcitedefaultmidpunct}
{\mcitedefaultendpunct}{\mcitedefaultseppunct}\relax
\EndOfBibitem
\bibitem[Mallory \emph{et~al.}(2014)Mallory, \ifmmode \check{S}\else
  \v{S}\fi{}ari\ifmmode~\acute{c}\else \'{c}\fi{}, Valeriani, and
  Cacciuto]{mall14}
S.~A. Mallory, A.~\ifmmode \check{S}\else \v{S}\fi{}ari\ifmmode~\acute{c}\else
  \'{c}\fi{}, C.~Valeriani and A.~Cacciuto, \emph{Phys. Rev. E}, 2014,
  \textbf{89}, 052303\relax
\mciteBstWouldAddEndPuncttrue
\mciteSetBstMidEndSepPunct{\mcitedefaultmidpunct}
{\mcitedefaultendpunct}{\mcitedefaultseppunct}\relax
\EndOfBibitem
\bibitem[Yang \emph{et~al.}(2014)Yang, Manning, and Marchetti]{yang14}
X.~Yang, M.~L. Manning and M.~C. Marchetti, \emph{Soft Matter}, 2014,
  \textbf{10}, 6477--6484\relax
\mciteBstWouldAddEndPuncttrue
\mciteSetBstMidEndSepPunct{\mcitedefaultmidpunct}
{\mcitedefaultendpunct}{\mcitedefaultseppunct}\relax
\EndOfBibitem
\bibitem[Takatori \emph{et~al.}(2014)Takatori, Yan, and Brady]{taka14}
S.~C. Takatori, W.~Yan and J.~F. Brady, \emph{Phys. Rev. Lett.}, 2014,
  \textbf{113}, 028103\relax
\mciteBstWouldAddEndPuncttrue
\mciteSetBstMidEndSepPunct{\mcitedefaultmidpunct}
{\mcitedefaultendpunct}{\mcitedefaultseppunct}\relax
\EndOfBibitem
\bibitem[Takatori and Brady(2014)]{taka14a}
S.~C. Takatori and J.~F. Brady, \emph{Soft Matter}, 2014, \textbf{10},
  9433--9445\relax
\mciteBstWouldAddEndPuncttrue
\mciteSetBstMidEndSepPunct{\mcitedefaultmidpunct}
{\mcitedefaultendpunct}{\mcitedefaultseppunct}\relax
\EndOfBibitem
\bibitem[Speck(2016)]{spec16}
T.~Speck, \emph{EPL}, 2016, \textbf{114}, 30006\relax
\mciteBstWouldAddEndPuncttrue
\mciteSetBstMidEndSepPunct{\mcitedefaultmidpunct}
{\mcitedefaultendpunct}{\mcitedefaultseppunct}\relax
\EndOfBibitem
\bibitem[Solon \emph{et~al.}(2015)Solon, Stenhammar, Wittkowski, Kardar, Kafri,
  Cates, and Tailleur]{solo15}
A.~P. Solon, J.~Stenhammar, R.~Wittkowski, M.~Kardar, Y.~Kafri, M.~E. Cates and
  J.~Tailleur, \emph{Phys. Rev. Lett.}, 2015, \textbf{114}, 198301\relax
\mciteBstWouldAddEndPuncttrue
\mciteSetBstMidEndSepPunct{\mcitedefaultmidpunct}
{\mcitedefaultendpunct}{\mcitedefaultseppunct}\relax
\EndOfBibitem
\bibitem[Winkler \emph{et~al.}(2015)Winkler, Wysocki, and Gompper]{wink15}
R.~G. Winkler, A.~Wysocki and G.~Gompper, \emph{Soft Matter}, 2015,
  \textbf{11}, 6680--6691\relax
\mciteBstWouldAddEndPuncttrue
\mciteSetBstMidEndSepPunct{\mcitedefaultmidpunct}
{\mcitedefaultendpunct}{\mcitedefaultseppunct}\relax
\EndOfBibitem
\bibitem[Levis \emph{et~al.}(2017)Levis, Codina, and Pagonabarraga]{levi17}
D.~Levis, J.~Codina and I.~Pagonabarraga, \emph{Soft Matter}, 2017,
  \textbf{13}, 8113--8119\relax
\mciteBstWouldAddEndPuncttrue
\mciteSetBstMidEndSepPunct{\mcitedefaultmidpunct}
{\mcitedefaultendpunct}{\mcitedefaultseppunct}\relax
\EndOfBibitem
\bibitem[Das \emph{et~al.}(2019)Das, Gompper, and Winkler]{das19}
S.~Das, G.~Gompper and R.~G. Winkler, \emph{Sci. Rep.}, 2019, \textbf{9},
  6608\relax
\mciteBstWouldAddEndPuncttrue
\mciteSetBstMidEndSepPunct{\mcitedefaultmidpunct}
{\mcitedefaultendpunct}{\mcitedefaultseppunct}\relax
\EndOfBibitem
\bibitem[Yan and Brady(2015)]{yan15a}
W.~Yan and J.~F. Brady, \emph{J. Fluid Mech.}, 2015, \textbf{785}, R1\relax
\mciteBstWouldAddEndPuncttrue
\mciteSetBstMidEndSepPunct{\mcitedefaultmidpunct}
{\mcitedefaultendpunct}{\mcitedefaultseppunct}\relax
\EndOfBibitem
\bibitem[Duzgun and Selinger(2018)]{duzgun18}
A.~Duzgun and J.~V. Selinger, \emph{Phys. Rev. E}, 2018, \textbf{97},
  032606\relax
\mciteBstWouldAddEndPuncttrue
\mciteSetBstMidEndSepPunct{\mcitedefaultmidpunct}
{\mcitedefaultendpunct}{\mcitedefaultseppunct}\relax
\EndOfBibitem
\bibitem[Ezhilan \emph{et~al.}(2015)Ezhilan, Alonso-Matilla, and
  Saintillan]{ezhi15}
B.~Ezhilan, R.~Alonso-Matilla and D.~Saintillan, \emph{J. Fluid Mech.}, 2015,
  \textbf{781}, R4\relax
\mciteBstWouldAddEndPuncttrue
\mciteSetBstMidEndSepPunct{\mcitedefaultmidpunct}
{\mcitedefaultendpunct}{\mcitedefaultseppunct}\relax
\EndOfBibitem
\bibitem[Ginot \emph{et~al.}(2015)Ginot, Theurkauff, Levis, Ybert, Bocquet,
  Berthier, and Cottin-Bizonne]{gino15}
F.~Ginot, I.~Theurkauff, D.~Levis, C.~Ybert, L.~Bocquet, L.~Berthier and
  C.~Cottin-Bizonne, \emph{Phys. Rev. X}, 2015, \textbf{5}, 011004\relax
\mciteBstWouldAddEndPuncttrue
\mciteSetBstMidEndSepPunct{\mcitedefaultmidpunct}
{\mcitedefaultendpunct}{\mcitedefaultseppunct}\relax
\EndOfBibitem
\bibitem[Junot \emph{et~al.}(2017)Junot, Briand, Ledesma-Alonso, and
  Dauchot]{juno17}
G.~Junot, G.~Briand, R.~Ledesma-Alonso and O.~Dauchot, \emph{Phys. Rev. Lett.},
  2017, \textbf{119}, 028002\relax
\mciteBstWouldAddEndPuncttrue
\mciteSetBstMidEndSepPunct{\mcitedefaultmidpunct}
{\mcitedefaultendpunct}{\mcitedefaultseppunct}\relax
\EndOfBibitem
\bibitem[Solon \emph{et~al.}(2015)Solon, Fily, Baskaran, Cates, Kafri, Kardar,
  and Tailleur]{solo15a}
A.~P. Solon, Y.~Fily, A.~Baskaran, M.~E. Cates, Y.~Kafri, M.~Kardar and
  J.~Tailleur, \emph{Nature Phys.}, 2015, \textbf{11}, 673--678\relax
\mciteBstWouldAddEndPuncttrue
\mciteSetBstMidEndSepPunct{\mcitedefaultmidpunct}
{\mcitedefaultendpunct}{\mcitedefaultseppunct}\relax
\EndOfBibitem
\bibitem[Omar \emph{et~al.}(2019)Omar, Wang, and Brady]{omar19}
A.~K. Omar, Z.-G. Wang and J.~F. Brady, \emph{arXiv:1912.11727}, 2019\relax
\mciteBstWouldAddEndPuncttrue
\mciteSetBstMidEndSepPunct{\mcitedefaultmidpunct}
{\mcitedefaultendpunct}{\mcitedefaultseppunct}\relax
\EndOfBibitem
\bibitem[Kirkwood and Buff(1949)]{kirk49}
J.~G. Kirkwood and F.~P. Buff, \emph{J. Chem. Phys.}, 1949, \textbf{17},
  338--343\relax
\mciteBstWouldAddEndPuncttrue
\mciteSetBstMidEndSepPunct{\mcitedefaultmidpunct}
{\mcitedefaultendpunct}{\mcitedefaultseppunct}\relax
\EndOfBibitem
\bibitem[Evans(1979)]{evan79}
R.~Evans, \emph{Adv. Phys.}, 1979, \textbf{28}, 143--200\relax
\mciteBstWouldAddEndPuncttrue
\mciteSetBstMidEndSepPunct{\mcitedefaultmidpunct}
{\mcitedefaultendpunct}{\mcitedefaultseppunct}\relax
\EndOfBibitem
\bibitem[Hohenberg and Krekhov(2015)]{hohe15}
P.~Hohenberg and A.~Krekhov, \emph{Phys. Rep.}, 2015, \textbf{572}, 1--42\relax
\mciteBstWouldAddEndPuncttrue
\mciteSetBstMidEndSepPunct{\mcitedefaultmidpunct}
{\mcitedefaultendpunct}{\mcitedefaultseppunct}\relax
\EndOfBibitem
\bibitem[Caussin \emph{et~al.}(2014)Caussin, Solon, Peshkov, Chat{\'{e}},
  Dauxois, Tailleur, Vitelli, and Bartolo]{caussin14}
J.-B. Caussin, A.~Solon, A.~Peshkov, H.~Chat{\'{e}}, T.~Dauxois, J.~Tailleur,
  V.~Vitelli and D.~Bartolo, \emph{Phys. Rev. Lett.}, 2014, \textbf{112},
  148102\relax
\mciteBstWouldAddEndPuncttrue
\mciteSetBstMidEndSepPunct{\mcitedefaultmidpunct}
{\mcitedefaultendpunct}{\mcitedefaultseppunct}\relax
\EndOfBibitem
\bibitem[Weitz \emph{et~al.}(2015)Weitz, Deutsch, and Peruani]{weitz15}
S.~Weitz, A.~Deutsch and F.~Peruani, \emph{Phys. Rev. E}, 2015, \textbf{92},
  012322\relax
\mciteBstWouldAddEndPuncttrue
\mciteSetBstMidEndSepPunct{\mcitedefaultmidpunct}
{\mcitedefaultendpunct}{\mcitedefaultseppunct}\relax
\EndOfBibitem
\bibitem[Bialk\'e \emph{et~al.}(2015)Bialk\'e, Siebert, L\"owen, and
  Speck]{bial15}
J.~Bialk\'e, J.~T. Siebert, H.~L\"owen and T.~Speck, \emph{Phys. Rev. Lett.},
  2015, \textbf{115}, 098301\relax
\mciteBstWouldAddEndPuncttrue
\mciteSetBstMidEndSepPunct{\mcitedefaultmidpunct}
{\mcitedefaultendpunct}{\mcitedefaultseppunct}\relax
\EndOfBibitem
\bibitem[Redner \emph{et~al.}(2013)Redner, Hagan, and Baskaran]{redn13}
G.~S. Redner, M.~F. Hagan and A.~Baskaran, \emph{Phys. Rev. Lett.}, 2013,
  \textbf{110}, 055701\relax
\mciteBstWouldAddEndPuncttrue
\mciteSetBstMidEndSepPunct{\mcitedefaultmidpunct}
{\mcitedefaultendpunct}{\mcitedefaultseppunct}\relax
\EndOfBibitem
\bibitem[Fily \emph{et~al.}(2014)Fily, Henkes, and Marchetti]{fily14}
Y.~Fily, S.~Henkes and M.~C. Marchetti, \emph{Soft Matter}, 2014, \textbf{10},
  2132--2140\relax
\mciteBstWouldAddEndPuncttrue
\mciteSetBstMidEndSepPunct{\mcitedefaultmidpunct}
{\mcitedefaultendpunct}{\mcitedefaultseppunct}\relax
\EndOfBibitem
\bibitem[Bialk\'e \emph{et~al.}(2015)Bialk\'e, Speck, and L\"owen]{bial14}
J.~Bialk\'e, T.~Speck and H.~L\"owen, \emph{J. Non-Cryst. Solids}, 2015,
  \textbf{407}, 367--375\relax
\mciteBstWouldAddEndPuncttrue
\mciteSetBstMidEndSepPunct{\mcitedefaultmidpunct}
{\mcitedefaultendpunct}{\mcitedefaultseppunct}\relax
\EndOfBibitem
\bibitem[Digregorio \emph{et~al.}(2018)Digregorio, Levis, Suma, Cugliandolo,
  Gonnella, and Pagonabarraga]{digregorio18}
P.~Digregorio, D.~Levis, A.~Suma, L.~F. Cugliandolo, G.~Gonnella and
  I.~Pagonabarraga, \emph{Phys. Rev. Lett.}, 2018, \textbf{121}, 098003\relax
\mciteBstWouldAddEndPuncttrue
\mciteSetBstMidEndSepPunct{\mcitedefaultmidpunct}
{\mcitedefaultendpunct}{\mcitedefaultseppunct}\relax
\EndOfBibitem
\bibitem[Siebert \emph{et~al.}(2018)Siebert, Dittrich, Schmid, Binder, Speck,
  and Virnau]{sieb18}
J.~T. Siebert, F.~Dittrich, F.~Schmid, K.~Binder, T.~Speck and P.~Virnau,
  \emph{Phys. Rev. E}, 2018, \textbf{98}, 030601\relax
\mciteBstWouldAddEndPuncttrue
\mciteSetBstMidEndSepPunct{\mcitedefaultmidpunct}
{\mcitedefaultendpunct}{\mcitedefaultseppunct}\relax
\EndOfBibitem
\bibitem[Binder(2003)]{bind03}
K.~Binder, \emph{Physica A}, 2003, \textbf{319}, 99--114\relax
\mciteBstWouldAddEndPuncttrue
\mciteSetBstMidEndSepPunct{\mcitedefaultmidpunct}
{\mcitedefaultendpunct}{\mcitedefaultseppunct}\relax
\EndOfBibitem
\bibitem[Schrader \emph{et~al.}(2009)Schrader, Virnau, and Binder]{schr09}
M.~Schrader, P.~Virnau and K.~Binder, \emph{Phys. Rev. E}, 2009, \textbf{79},
  061104\relax
\mciteBstWouldAddEndPuncttrue
\mciteSetBstMidEndSepPunct{\mcitedefaultmidpunct}
{\mcitedefaultendpunct}{\mcitedefaultseppunct}\relax
\EndOfBibitem
\bibitem[Paliwal \emph{et~al.}(2017)Paliwal, Prymidis, Filion, and
  Dijkstra]{paliwal17}
S.~Paliwal, V.~Prymidis, L.~Filion and M.~Dijkstra, \emph{J. Chem. Phys.},
  2017, \textbf{147}, 084902\relax
\mciteBstWouldAddEndPuncttrue
\mciteSetBstMidEndSepPunct{\mcitedefaultmidpunct}
{\mcitedefaultendpunct}{\mcitedefaultseppunct}\relax
\EndOfBibitem
\bibitem[Solon \emph{et~al.}(2018)Solon, Stenhammar, Cates, Kafri, and
  Tailleur]{solo17}
A.~P. Solon, J.~Stenhammar, M.~E. Cates, Y.~Kafri and J.~Tailleur, \emph{Phys.
  Rev. E}, 2018, \textbf{97}, 020602\relax
\mciteBstWouldAddEndPuncttrue
\mciteSetBstMidEndSepPunct{\mcitedefaultmidpunct}
{\mcitedefaultendpunct}{\mcitedefaultseppunct}\relax
\EndOfBibitem
\bibitem[Paliwal \emph{et~al.}(2018)Paliwal, Rodenburg, van Roij, and
  Dijkstra]{paliwal18}
S.~Paliwal, J.~Rodenburg, R.~van Roij and M.~Dijkstra, \emph{New J. Phys.},
  2018, \textbf{20}, 015003\relax
\mciteBstWouldAddEndPuncttrue
\mciteSetBstMidEndSepPunct{\mcitedefaultmidpunct}
{\mcitedefaultendpunct}{\mcitedefaultseppunct}\relax
\EndOfBibitem
\bibitem[Hermann \emph{et~al.}(2019)Hermann, Krinninger, de~las Heras, and
  Schmidt]{herm19a}
S.~Hermann, P.~Krinninger, D.~de~las Heras and M.~Schmidt, \emph{Phys. Rev. E},
  2019, \textbf{100}, 052604\relax
\mciteBstWouldAddEndPuncttrue
\mciteSetBstMidEndSepPunct{\mcitedefaultmidpunct}
{\mcitedefaultendpunct}{\mcitedefaultseppunct}\relax
\EndOfBibitem
\bibitem[Hermann \emph{et~al.}(2019)Hermann, de~las Heras, and
  Schmidt]{herm19b}
S.~Hermann, D.~de~las Heras and M.~Schmidt, \emph{Phys. Rev. Lett.}, 2019,
  \textbf{123}, 268002\relax
\mciteBstWouldAddEndPuncttrue
\mciteSetBstMidEndSepPunct{\mcitedefaultmidpunct}
{\mcitedefaultendpunct}{\mcitedefaultseppunct}\relax
\EndOfBibitem
\bibitem[Wittmann \emph{et~al.}(2019)Wittmann, Smallenburg, and
  Brader]{wittmann19}
R.~Wittmann, F.~Smallenburg and J.~M. Brader, \emph{J. Chem. Phys.}, 2019,
  \textbf{150}, 174908\relax
\mciteBstWouldAddEndPuncttrue
\mciteSetBstMidEndSepPunct{\mcitedefaultmidpunct}
{\mcitedefaultendpunct}{\mcitedefaultseppunct}\relax
\EndOfBibitem
\bibitem[Hohenberg and Halperin(1977)]{hohe77}
P.~C. Hohenberg and B.~I. Halperin, \emph{Rev. Mod. Phys.}, 1977, \textbf{49},
  435--479\relax
\mciteBstWouldAddEndPuncttrue
\mciteSetBstMidEndSepPunct{\mcitedefaultmidpunct}
{\mcitedefaultendpunct}{\mcitedefaultseppunct}\relax
\EndOfBibitem
\bibitem[Speck \emph{et~al.}(2014)Speck, Bialk\'e, Menzel, and L\"owen]{spec14}
T.~Speck, J.~Bialk\'e, A.~M. Menzel and H.~L\"owen, \emph{Phys. Rev. Lett.},
  2014, \textbf{112}, 218304\relax
\mciteBstWouldAddEndPuncttrue
\mciteSetBstMidEndSepPunct{\mcitedefaultmidpunct}
{\mcitedefaultendpunct}{\mcitedefaultseppunct}\relax
\EndOfBibitem
\bibitem[Tjhung \emph{et~al.}(2018)Tjhung, Nardini, and Cates]{tjhu18}
E.~Tjhung, C.~Nardini and M.~E. Cates, \emph{Phys. Rev. X}, 2018, \textbf{8},
  031080\relax
\mciteBstWouldAddEndPuncttrue
\mciteSetBstMidEndSepPunct{\mcitedefaultmidpunct}
{\mcitedefaultendpunct}{\mcitedefaultseppunct}\relax
\EndOfBibitem
\bibitem[Wittkowski \emph{et~al.}(2014)Wittkowski, Tiribocchi, Stenhammar,
  Allen, Marenduzzo, and Cates]{witt14}
R.~Wittkowski, A.~Tiribocchi, J.~Stenhammar, R.~J. Allen, D.~Marenduzzo and
  M.~E. Cates, \emph{Nat. Commun.}, 2014, \textbf{5}, 4351\relax
\mciteBstWouldAddEndPuncttrue
\mciteSetBstMidEndSepPunct{\mcitedefaultmidpunct}
{\mcitedefaultendpunct}{\mcitedefaultseppunct}\relax
\EndOfBibitem
\bibitem[Bickmann and Wittkowski(2019)]{bickmann19}
J.~Bickmann and R.~Wittkowski, \emph{J. Phys. Condens. Matter}, 2019\relax
\mciteBstWouldAddEndPuncttrue
\mciteSetBstMidEndSepPunct{\mcitedefaultmidpunct}
{\mcitedefaultendpunct}{\mcitedefaultseppunct}\relax
\EndOfBibitem
\bibitem[Rapp \emph{et~al.}(2019)Rapp, Bergmann, and Zimmermann]{rapp19}
L.~Rapp, F.~Bergmann and W.~Zimmermann, \emph{Eur. Phys. J. E}, 2019,
  \textbf{42}, 57\relax
\mciteBstWouldAddEndPuncttrue
\mciteSetBstMidEndSepPunct{\mcitedefaultmidpunct}
{\mcitedefaultendpunct}{\mcitedefaultseppunct}\relax
\EndOfBibitem
\bibitem[Palacci \emph{et~al.}(2013)Palacci, Sacanna, Vatchinsky, Chaikin, and
  Pine]{pala13a}
J.~Palacci, S.~Sacanna, A.~Vatchinsky, P.~M. Chaikin and D.~J. Pine, \emph{J.
  Am. Chem. Soc.}, 2013, \textbf{135}, 15978--15981\relax
\mciteBstWouldAddEndPuncttrue
\mciteSetBstMidEndSepPunct{\mcitedefaultmidpunct}
{\mcitedefaultendpunct}{\mcitedefaultseppunct}\relax
\EndOfBibitem
\bibitem[Lozano \emph{et~al.}(2016)Lozano, ten Hagen, L\"owen, and
  Bechinger]{loza16}
C.~Lozano, B.~ten Hagen, H.~L\"owen and C.~Bechinger, \emph{Nat. Commun.},
  2016, \textbf{7}, 12828\relax
\mciteBstWouldAddEndPuncttrue
\mciteSetBstMidEndSepPunct{\mcitedefaultmidpunct}
{\mcitedefaultendpunct}{\mcitedefaultseppunct}\relax
\EndOfBibitem
\bibitem[Fischer \emph{et~al.}(2020)Fischer, Schmid, and Speck]{fisc20}
A.~Fischer, F.~Schmid and T.~Speck, \emph{Phys. Rev. E}, 2020, \textbf{101},
  012601\relax
\mciteBstWouldAddEndPuncttrue
\mciteSetBstMidEndSepPunct{\mcitedefaultmidpunct}
{\mcitedefaultendpunct}{\mcitedefaultseppunct}\relax
\EndOfBibitem
\bibitem[Galajda \emph{et~al.}(2007)Galajda, Keymer, Chaikin, and
  Austin]{galajda07}
P.~Galajda, J.~Keymer, P.~Chaikin and R.~Austin, \emph{J. Bacteriol.}, 2007,
  \textbf{189}, 8704--8707\relax
\mciteBstWouldAddEndPuncttrue
\mciteSetBstMidEndSepPunct{\mcitedefaultmidpunct}
{\mcitedefaultendpunct}{\mcitedefaultseppunct}\relax
\EndOfBibitem
\bibitem[Baek \emph{et~al.}(2018)Baek, Solon, Xu, Nikola, and Kafri]{baek18}
Y.~Baek, A.~P. Solon, X.~Xu, N.~Nikola and Y.~Kafri, \emph{Phys. Rev. Lett.},
  2018, \textbf{120}, 058002\relax
\mciteBstWouldAddEndPuncttrue
\mciteSetBstMidEndSepPunct{\mcitedefaultmidpunct}
{\mcitedefaultendpunct}{\mcitedefaultseppunct}\relax
\EndOfBibitem
\bibitem[Kaiser \emph{et~al.}(2012)Kaiser, Wensink, and L\"owen]{kais12}
A.~Kaiser, H.~H. Wensink and H.~L\"owen, \emph{Phys. Rev. Lett.}, 2012,
  \textbf{108}, 268307\relax
\mciteBstWouldAddEndPuncttrue
\mciteSetBstMidEndSepPunct{\mcitedefaultmidpunct}
{\mcitedefaultendpunct}{\mcitedefaultseppunct}\relax
\EndOfBibitem
\bibitem[Wan \emph{et~al.}(2008)Wan, Olson~Reichhardt, Nussinov, and
  Reichhardt]{wan08}
M.~B. Wan, C.~J. Olson~Reichhardt, Z.~Nussinov and C.~Reichhardt, \emph{Phys.
  Rev. Lett.}, 2008, \textbf{101}, 018102\relax
\mciteBstWouldAddEndPuncttrue
\mciteSetBstMidEndSepPunct{\mcitedefaultmidpunct}
{\mcitedefaultendpunct}{\mcitedefaultseppunct}\relax
\EndOfBibitem
\bibitem[Stenhammar \emph{et~al.}(2016)Stenhammar, Wittkowski, Marenduzzo, and
  Cates]{sten16}
J.~Stenhammar, R.~Wittkowski, D.~Marenduzzo and M.~E. Cates, \emph{Sci. Adv.},
  2016, \textbf{2}, e1501850\relax
\mciteBstWouldAddEndPuncttrue
\mciteSetBstMidEndSepPunct{\mcitedefaultmidpunct}
{\mcitedefaultendpunct}{\mcitedefaultseppunct}\relax
\EndOfBibitem
\bibitem[Ni \emph{et~al.}(2015)Ni, Cohen~Stuart, and Bolhuis]{ni15}
R.~Ni, M.~A. Cohen~Stuart and P.~G. Bolhuis, \emph{Phys. Rev. Lett.}, 2015,
  \textbf{114}, 018302\relax
\mciteBstWouldAddEndPuncttrue
\mciteSetBstMidEndSepPunct{\mcitedefaultmidpunct}
{\mcitedefaultendpunct}{\mcitedefaultseppunct}\relax
\EndOfBibitem
\bibitem[Yamchi and Naji(2017)]{yamchi17}
M.~Z. Yamchi and A.~Naji, \emph{J. Chem. Phys.}, 2017, \textbf{147},
  194901\relax
\mciteBstWouldAddEndPuncttrue
\mciteSetBstMidEndSepPunct{\mcitedefaultmidpunct}
{\mcitedefaultendpunct}{\mcitedefaultseppunct}\relax
\EndOfBibitem
\bibitem[Harder \emph{et~al.}(2014)Harder, Mallory, Tung, Valeriani, and
  Cacciuto]{harder14}
J.~Harder, S.~A. Mallory, C.~Tung, C.~Valeriani and A.~Cacciuto, \emph{J. Chem.
  Phys.}, 2014, \textbf{141}, 194901\relax
\mciteBstWouldAddEndPuncttrue
\mciteSetBstMidEndSepPunct{\mcitedefaultmidpunct}
{\mcitedefaultendpunct}{\mcitedefaultseppunct}\relax
\EndOfBibitem
\bibitem[Nikola \emph{et~al.}(2016)Nikola, Solon, Kafri, Kardar, Tailleur, and
  Voituriez]{niko16}
N.~Nikola, A.~P. Solon, Y.~Kafri, M.~Kardar, J.~Tailleur and R.~Voituriez,
  \emph{Phys. Rev. Lett.}, 2016, \textbf{117}, 098001\relax
\mciteBstWouldAddEndPuncttrue
\mciteSetBstMidEndSepPunct{\mcitedefaultmidpunct}
{\mcitedefaultendpunct}{\mcitedefaultseppunct}\relax
\EndOfBibitem
\bibitem[Ivlev \emph{et~al.}(2015)Ivlev, Bartnick, Heinen, Du, Nosenko, and
  L\"owen]{ivlev15}
A.~V. Ivlev, J.~Bartnick, M.~Heinen, C.-R. Du, V.~Nosenko and H.~L\"owen,
  \emph{Phys. Rev. X}, 2015, \textbf{5}, 011035\relax
\mciteBstWouldAddEndPuncttrue
\mciteSetBstMidEndSepPunct{\mcitedefaultmidpunct}
{\mcitedefaultendpunct}{\mcitedefaultseppunct}\relax
\EndOfBibitem
\bibitem[Shin \emph{et~al.}(2017)Shin, Cherstvy, Kim, and Zaburdaev]{shin17}
J.~Shin, A.~G. Cherstvy, W.~K. Kim and V.~Zaburdaev, \emph{Phys. Chem. Chem.
  Phys.}, 2017, \textbf{19}, 18338--18347\relax
\mciteBstWouldAddEndPuncttrue
\mciteSetBstMidEndSepPunct{\mcitedefaultmidpunct}
{\mcitedefaultendpunct}{\mcitedefaultseppunct}\relax
\EndOfBibitem
\bibitem[Mallory \emph{et~al.}(2015)Mallory, Valeriani, and
  Cacciuto]{mallory15}
S.~A. Mallory, C.~Valeriani and A.~Cacciuto, \emph{Phys. Rev. E}, 2015,
  \textbf{92}, 012314\relax
\mciteBstWouldAddEndPuncttrue
\mciteSetBstMidEndSepPunct{\mcitedefaultmidpunct}
{\mcitedefaultendpunct}{\mcitedefaultseppunct}\relax
\EndOfBibitem
\bibitem[Ray \emph{et~al.}(2014)Ray, Reichhardt, and Reichhardt]{ray14}
D.~Ray, C.~Reichhardt and C.~J.~O. Reichhardt, \emph{Phys. Rev. E}, 2014,
  \textbf{90}, 013019\relax
\mciteBstWouldAddEndPuncttrue
\mciteSetBstMidEndSepPunct{\mcitedefaultmidpunct}
{\mcitedefaultendpunct}{\mcitedefaultseppunct}\relax
\EndOfBibitem
\bibitem[Rohwer \emph{et~al.}(2017)Rohwer, Kardar, and Kr\"uger]{rohwer17}
C.~M. Rohwer, M.~Kardar and M.~Kr\"uger, \emph{Phys. Rev. Lett.}, 2017,
  \textbf{118}, 015702\relax
\mciteBstWouldAddEndPuncttrue
\mciteSetBstMidEndSepPunct{\mcitedefaultmidpunct}
{\mcitedefaultendpunct}{\mcitedefaultseppunct}\relax
\EndOfBibitem
\bibitem[Seifert(2012)]{seif12}
U.~Seifert, \emph{Rep. Prog. Phys.}, 2012, \textbf{75}, 126001\relax
\mciteBstWouldAddEndPuncttrue
\mciteSetBstMidEndSepPunct{\mcitedefaultmidpunct}
{\mcitedefaultendpunct}{\mcitedefaultseppunct}\relax
\EndOfBibitem
\bibitem[Seifert(2019)]{seif19}
U.~Seifert, \emph{Annu. Rev. Condens. Matter Phys.}, 2019, \textbf{10},
  171--192\relax
\mciteBstWouldAddEndPuncttrue
\mciteSetBstMidEndSepPunct{\mcitedefaultmidpunct}
{\mcitedefaultendpunct}{\mcitedefaultseppunct}\relax
\EndOfBibitem
\bibitem[Fodor \emph{et~al.}(2016)Fodor, Nardini, Cates, Tailleur, Visco, and
  van Wijland]{fodo16}
E.~Fodor, C.~Nardini, M.~E. Cates, J.~Tailleur, P.~Visco and F.~van Wijland,
  \emph{Phys. Rev. Lett.}, 2016, \textbf{117}, 038103\relax
\mciteBstWouldAddEndPuncttrue
\mciteSetBstMidEndSepPunct{\mcitedefaultmidpunct}
{\mcitedefaultendpunct}{\mcitedefaultseppunct}\relax
\EndOfBibitem
\bibitem[Mandal \emph{et~al.}(2017)Mandal, Klymko, and DeWeese]{mand17}
D.~Mandal, K.~Klymko and M.~R. DeWeese, \emph{Phys. Rev. Lett.}, 2017,
  \textbf{119}, 258001\relax
\mciteBstWouldAddEndPuncttrue
\mciteSetBstMidEndSepPunct{\mcitedefaultmidpunct}
{\mcitedefaultendpunct}{\mcitedefaultseppunct}\relax
\EndOfBibitem
\bibitem[Caprini \emph{et~al.}(2018)Caprini, Marconi, Puglisi, and
  Vulpiani]{caprini18}
L.~Caprini, U.~M.~B. Marconi, A.~Puglisi and A.~Vulpiani, \emph{Phys. Rev.
  Lett.}, 2018, \textbf{121}, 139801\relax
\mciteBstWouldAddEndPuncttrue
\mciteSetBstMidEndSepPunct{\mcitedefaultmidpunct}
{\mcitedefaultendpunct}{\mcitedefaultseppunct}\relax
\EndOfBibitem
\bibitem[Dabelow \emph{et~al.}(2019)Dabelow, Bo, and Eichhorn]{dabe19}
L.~Dabelow, S.~Bo and R.~Eichhorn, \emph{Phys. Rev. X}, 2019, \textbf{9},
  021009\relax
\mciteBstWouldAddEndPuncttrue
\mciteSetBstMidEndSepPunct{\mcitedefaultmidpunct}
{\mcitedefaultendpunct}{\mcitedefaultseppunct}\relax
\EndOfBibitem
\bibitem[Nardini \emph{et~al.}(2017)Nardini, Fodor, Tjhung, van Wijland,
  Tailleur, and Cates]{nard17}
C.~Nardini, E.~Fodor, E.~Tjhung, F.~van Wijland, J.~Tailleur and M.~E. Cates,
  \emph{Phys. Rev. X}, 2017, \textbf{7}, 021007\relax
\mciteBstWouldAddEndPuncttrue
\mciteSetBstMidEndSepPunct{\mcitedefaultmidpunct}
{\mcitedefaultendpunct}{\mcitedefaultseppunct}\relax
\EndOfBibitem
\bibitem[Ganguly and Chaudhuri(2013)]{gang13}
C.~Ganguly and D.~Chaudhuri, \emph{Phys. Rev. E}, 2013, \textbf{88},
  032102\relax
\mciteBstWouldAddEndPuncttrue
\mciteSetBstMidEndSepPunct{\mcitedefaultmidpunct}
{\mcitedefaultendpunct}{\mcitedefaultseppunct}\relax
\EndOfBibitem
\bibitem[Falasco \emph{et~al.}(2016)Falasco, Pfaller, Bregulla, Cichos, and
  Kroy]{fala16}
G.~Falasco, R.~Pfaller, A.~P. Bregulla, F.~Cichos and K.~Kroy, \emph{Phys. Rev.
  E}, 2016, \textbf{94}, 030602\relax
\mciteBstWouldAddEndPuncttrue
\mciteSetBstMidEndSepPunct{\mcitedefaultmidpunct}
{\mcitedefaultendpunct}{\mcitedefaultseppunct}\relax
\EndOfBibitem
\bibitem[Marconi \emph{et~al.}(2017)Marconi, Puglisi, and Maggi]{marc17}
U.~M.~B. Marconi, A.~Puglisi and C.~Maggi, \emph{Sci. Rep.}, 2017, \textbf{7},
  46496\relax
\mciteBstWouldAddEndPuncttrue
\mciteSetBstMidEndSepPunct{\mcitedefaultmidpunct}
{\mcitedefaultendpunct}{\mcitedefaultseppunct}\relax
\EndOfBibitem
\bibitem[Puglisi and Marconi(2017)]{puglisi17}
A.~Puglisi and U.~M.~B. Marconi, \emph{Entropy}, 2017, \textbf{19}, 356\relax
\mciteBstWouldAddEndPuncttrue
\mciteSetBstMidEndSepPunct{\mcitedefaultmidpunct}
{\mcitedefaultendpunct}{\mcitedefaultseppunct}\relax
\EndOfBibitem
\bibitem[Shankar and Marchetti(2018)]{shankar18}
S.~Shankar and M.~C. Marchetti, \emph{Phys. Rev. E}, 2018, \textbf{98},
  020604\relax
\mciteBstWouldAddEndPuncttrue
\mciteSetBstMidEndSepPunct{\mcitedefaultmidpunct}
{\mcitedefaultendpunct}{\mcitedefaultseppunct}\relax
\EndOfBibitem
\bibitem[Crosato \emph{et~al.}(2019)Crosato, Prokopenko, and
  Spinney]{crosato19}
E.~Crosato, M.~Prokopenko and R.~E. Spinney, \emph{Phys. Rev. E}, 2019,
  \textbf{100}, 042613\relax
\mciteBstWouldAddEndPuncttrue
\mciteSetBstMidEndSepPunct{\mcitedefaultmidpunct}
{\mcitedefaultendpunct}{\mcitedefaultseppunct}\relax
\EndOfBibitem
\bibitem[Note1()]{Note1}
Note that $\Delta \mu $ relates to the chemical potential difference of two
  chemical species and is not to be confused with the effective chemical
  potential in Sec.~\ref {sec:coex}.\relax
\mciteBstWouldAddEndPunctfalse
\mciteSetBstMidEndSepPunct{\mcitedefaultmidpunct}
{}{\mcitedefaultseppunct}\relax
\EndOfBibitem
\bibitem[Gaspard and Kapral(2017)]{gasp17}
P.~Gaspard and R.~Kapral, \emph{J. Chem. Phys.}, 2017, \textbf{147},
  211101\relax
\mciteBstWouldAddEndPuncttrue
\mciteSetBstMidEndSepPunct{\mcitedefaultmidpunct}
{\mcitedefaultendpunct}{\mcitedefaultseppunct}\relax
\EndOfBibitem
\bibitem[Speck(2018)]{spec18}
T.~Speck, \emph{{EPL} (Europhysics Letters)}, 2018, \textbf{123}, 20007\relax
\mciteBstWouldAddEndPuncttrue
\mciteSetBstMidEndSepPunct{\mcitedefaultmidpunct}
{\mcitedefaultendpunct}{\mcitedefaultseppunct}\relax
\EndOfBibitem
\bibitem[Pietzonka and Seifert(2017)]{piet17}
P.~Pietzonka and U.~Seifert, \emph{J. Phys. A: Math. Theor.}, 2017,
  \textbf{51}, 01LT01\relax
\mciteBstWouldAddEndPuncttrue
\mciteSetBstMidEndSepPunct{\mcitedefaultmidpunct}
{\mcitedefaultendpunct}{\mcitedefaultseppunct}\relax
\EndOfBibitem
\bibitem[Pietzonka \emph{et~al.}(2019)Pietzonka, Fodor, Lohrmann, Cates, and
  Seifert]{piet19}
P.~Pietzonka, E.~Fodor, C.~Lohrmann, M.~E. Cates and U.~Seifert, \emph{Phys.
  Rev. X}, 2019, \textbf{9}, 041032\relax
\mciteBstWouldAddEndPuncttrue
\mciteSetBstMidEndSepPunct{\mcitedefaultmidpunct}
{\mcitedefaultendpunct}{\mcitedefaultseppunct}\relax
\EndOfBibitem
\bibitem[Krishnamurthy \emph{et~al.}(2016)Krishnamurthy, Ghosh, Chatterji,
  Ganapathy, and Sood]{krishna16}
S.~Krishnamurthy, S.~Ghosh, D.~Chatterji, R.~Ganapathy and A.~K. Sood,
  \emph{Nat. Phys.}, 2016, \textbf{12}, 1134--1138\relax
\mciteBstWouldAddEndPuncttrue
\mciteSetBstMidEndSepPunct{\mcitedefaultmidpunct}
{\mcitedefaultendpunct}{\mcitedefaultseppunct}\relax
\EndOfBibitem
\bibitem[Vizsnyiczai \emph{et~al.}(2017)Vizsnyiczai, Frangipane, Maggi,
  Saglimbeni, Bianchi, and Leonardo]{vizs17}
G.~Vizsnyiczai, G.~Frangipane, C.~Maggi, F.~Saglimbeni, S.~Bianchi and R.~D.
  Leonardo, \emph{Nat. Commun.}, 2017, \textbf{8}, 15974\relax
\mciteBstWouldAddEndPuncttrue
\mciteSetBstMidEndSepPunct{\mcitedefaultmidpunct}
{\mcitedefaultendpunct}{\mcitedefaultseppunct}\relax
\EndOfBibitem
\bibitem[Souslov \emph{et~al.}(2017)Souslov, van Zuiden, Bartolo, and
  Vitelli]{souslov17}
A.~Souslov, B.~C. van Zuiden, D.~Bartolo and V.~Vitelli, \emph{Nat. Phys.},
  2017, \textbf{13}, 1091--1094\relax
\mciteBstWouldAddEndPuncttrue
\mciteSetBstMidEndSepPunct{\mcitedefaultmidpunct}
{\mcitedefaultendpunct}{\mcitedefaultseppunct}\relax
\EndOfBibitem
\bibitem[Whitelam \emph{et~al.}(2009)Whitelam, Feng, Hagan, and
  Geissler]{whit09}
S.~Whitelam, E.~H. Feng, M.~F. Hagan and P.~L. Geissler, \emph{Soft Matter},
  2009, \textbf{5}, 1251--1262\relax
\mciteBstWouldAddEndPuncttrue
\mciteSetBstMidEndSepPunct{\mcitedefaultmidpunct}
{\mcitedefaultendpunct}{\mcitedefaultseppunct}\relax
\EndOfBibitem
\bibitem[K\"ummel \emph{et~al.}(2015)K\"ummel, Shabestari, Lozano, Volpe, and
  Bechinger]{kumm15}
F.~K\"ummel, P.~Shabestari, C.~Lozano, G.~Volpe and C.~Bechinger, \emph{Soft
  Matter}, 2015, \textbf{11}, 6187--6191\relax
\mciteBstWouldAddEndPuncttrue
\mciteSetBstMidEndSepPunct{\mcitedefaultmidpunct}
{\mcitedefaultendpunct}{\mcitedefaultseppunct}\relax
\EndOfBibitem
\bibitem[van~der Meer \emph{et~al.}(2016)van~der Meer, Dijkstra, and
  Filion]{meer16}
B.~van~der Meer, M.~Dijkstra and L.~Filion, \emph{Soft Matter}, 2016,
  \textbf{12}, 5630--5635\relax
\mciteBstWouldAddEndPuncttrue
\mciteSetBstMidEndSepPunct{\mcitedefaultmidpunct}
{\mcitedefaultendpunct}{\mcitedefaultseppunct}\relax
\EndOfBibitem
\bibitem[Mallory and Cacciuto(2019)]{mallory19}
S.~A. Mallory and A.~Cacciuto, \emph{Journal of the American Chemical Society},
  2019, \textbf{141}, 2500--2507\relax
\mciteBstWouldAddEndPuncttrue
\mciteSetBstMidEndSepPunct{\mcitedefaultmidpunct}
{\mcitedefaultendpunct}{\mcitedefaultseppunct}\relax
\EndOfBibitem
\bibitem[Wykes \emph{et~al.}(2016)Wykes, Palacci, Adachi, Ristroph, Zhong,
  Ward, Zhang, and Shelley]{wykes16}
M.~S.~D. Wykes, J.~Palacci, T.~Adachi, L.~Ristroph, X.~Zhong, M.~D. Ward,
  J.~Zhang and M.~J. Shelley, \emph{Soft Matter}, 2016, \textbf{12},
  4584--4589\relax
\mciteBstWouldAddEndPuncttrue
\mciteSetBstMidEndSepPunct{\mcitedefaultmidpunct}
{\mcitedefaultendpunct}{\mcitedefaultseppunct}\relax
\EndOfBibitem
\bibitem[Niu and Palberg(2018)]{niu18}
R.~Niu and T.~Palberg, \emph{Soft Matter}, 2018, \textbf{14}, 7554--7568\relax
\mciteBstWouldAddEndPuncttrue
\mciteSetBstMidEndSepPunct{\mcitedefaultmidpunct}
{\mcitedefaultendpunct}{\mcitedefaultseppunct}\relax
\EndOfBibitem
\bibitem[Xie \emph{et~al.}(2019)Xie, Sun, Fan, Lin, Chen, Wang, Dong, and
  He]{xie19}
H.~Xie, M.~Sun, X.~Fan, Z.~Lin, W.~Chen, L.~Wang, L.~Dong and Q.~He, \emph{Sci.
  Robot.}, 2019, \textbf{4}, eaav8006\relax
\mciteBstWouldAddEndPuncttrue
\mciteSetBstMidEndSepPunct{\mcitedefaultmidpunct}
{\mcitedefaultendpunct}{\mcitedefaultseppunct}\relax
\EndOfBibitem
\bibitem[Li \emph{et~al.}(2019)Li, Batra, Brown, Chang, Ranganathan, Hoberman,
  Rus, and Lipson]{li19}
S.~Li, R.~Batra, D.~Brown, H.-D. Chang, N.~Ranganathan, C.~Hoberman, D.~Rus and
  H.~Lipson, \emph{Nature}, 2019, \textbf{567}, 361--365\relax
\mciteBstWouldAddEndPuncttrue
\mciteSetBstMidEndSepPunct{\mcitedefaultmidpunct}
{\mcitedefaultendpunct}{\mcitedefaultseppunct}\relax
\EndOfBibitem
\bibitem[Brangwynne \emph{et~al.}(2015)Brangwynne, Tompa, and Pappu]{bran15}
C.~P. Brangwynne, P.~Tompa and R.~V. Pappu, \emph{Nat. Phys.}, 2015,
  \textbf{11}, 899--904\relax
\mciteBstWouldAddEndPuncttrue
\mciteSetBstMidEndSepPunct{\mcitedefaultmidpunct}
{\mcitedefaultendpunct}{\mcitedefaultseppunct}\relax
\EndOfBibitem
\bibitem[Brangwynne \emph{et~al.}(2009)Brangwynne, Eckmann, Courson, Rybarska,
  Hoege, Gharakhani, Julicher, and Hyman]{bran09}
C.~P. Brangwynne, C.~R. Eckmann, D.~S. Courson, A.~Rybarska, C.~Hoege,
  J.~Gharakhani, F.~Julicher and A.~A. Hyman, \emph{Science}, 2009,
  \textbf{324}, 1729--1732\relax
\mciteBstWouldAddEndPuncttrue
\mciteSetBstMidEndSepPunct{\mcitedefaultmidpunct}
{\mcitedefaultendpunct}{\mcitedefaultseppunct}\relax
\EndOfBibitem
\bibitem[Hyman \emph{et~al.}(2014)Hyman, Weber, and Jülicher]{hyman14}
A.~A. Hyman, C.~A. Weber and F.~Jülicher, \emph{Annu. Rev. Cell Dev. Biol.},
  2014, \textbf{30}, 39--58\relax
\mciteBstWouldAddEndPuncttrue
\mciteSetBstMidEndSepPunct{\mcitedefaultmidpunct}
{\mcitedefaultendpunct}{\mcitedefaultseppunct}\relax
\EndOfBibitem
\bibitem[Sabass \emph{et~al.}(2008)Sabass, Gardel, Waterman, and
  Schwarz]{sabass08}
B.~Sabass, M.~L. Gardel, C.~M. Waterman and U.~S. Schwarz, \emph{Biophys. J.},
  2008, \textbf{94}, 207--220\relax
\mciteBstWouldAddEndPuncttrue
\mciteSetBstMidEndSepPunct{\mcitedefaultmidpunct}
{\mcitedefaultendpunct}{\mcitedefaultseppunct}\relax
\EndOfBibitem
\bibitem[Trepat \emph{et~al.}(2009)Trepat, Wasserman, Angelini, Millet, Weitz,
  Butler, and Fredberg]{trepat09}
X.~Trepat, M.~R. Wasserman, T.~E. Angelini, E.~Millet, D.~A. Weitz, J.~P.
  Butler and J.~J. Fredberg, \emph{Nat. Phys.}, 2009, \textbf{5},
  426--430\relax
\mciteBstWouldAddEndPuncttrue
\mciteSetBstMidEndSepPunct{\mcitedefaultmidpunct}
{\mcitedefaultendpunct}{\mcitedefaultseppunct}\relax
\EndOfBibitem
\end{mcitethebibliography}
\end{document}